\documentclass[journal]{IEEEtran}
\ifCLASSINFOpdf
\else
\fi
\usepackage{graphicx}
\usepackage{caption}
\usepackage{acro}
\usepackage{booktabs}
\acsetup{first-style=short}

\DeclareAcronym{sa}{
  short = SA ,
  long  = self-awareness ,
  class = abbrev
}
\DeclareAcronym{pl}{
  short = PL ,
  long  = private layer ,
  class = abbrev
}
\DeclareAcronym{sl}{
  short = SL ,
  long  = shared layer ,
  class = abbrev
}
\DeclareAcronym{sds}{
  short = SDS ,
  long  = switching dynamical system,
  class = abbrev
}
\DeclareAcronym{dbn2}{
  short = DBN ,
  long  = dynamical Bayesian network,
  class = abbrev
}

\DeclareAcronym{dbn}{
  short = DBN ,
  long  = dynamical Bayesian network,
  class = abbrev
}

\DeclareAcronym{kf}{
  short = KF ,
  long  = Kalman filter,
  class = abbrev
}
\DeclareAcronym{pf}{
  short = PF ,
  long  = particle filter,
  class = abbrev
}
\DeclareAcronym{gan}{
  short = GAN ,
  long  = generative adversarial networks,
  class = abbrev
}

\DeclareAcronym{umkf}{
  short = U-KF ,
  long  = unmotivated-agent Kalman filter,
  class = abbrev
}

\DeclareAcronym{mkf}{
  short = MKF ,
  long  = motivated Kalman filter,
  class = abbrev
}

\DeclareAcronym{hmm}{
  short = HMM ,
  long  = hidden Markov model,
  class = abbrev
}

\DeclareAcronym{mjpf}{
  short = MJPF ,
  long  = Markov jump particle filter,
  class = abbrev
}

\DeclareAcronym{som}{
  short = SOM ,
  long  = self-organizing map,
  class = abbrev
}

\DeclareAcronym{angelsperarea}{
  short = \ensuremath{a} ,
  long  = The number of angels per unit area ,
  sort  = a ,
  class = nomencl
}
\DeclareAcronym{numofangels}{
  short = \ensuremath{N} ,
  long  = The number of angels per needle point ,
  sort  = N ,
  class = nomencl
}
\DeclareAcronym{areaofneedle}{
  short = \ensuremath{A} ,
  long  = The area of the needle point ,
  sort  = A ,
  class = nomencl
}

\usepackage{color}
\usepackage{float}
\usepackage{amsmath,amsfonts,amssymb,bm}

\DeclareMathOperator*{\argmin}{arg\,min}

\usepackage{adjustbox}
\usepackage[noadjust]{cite}
\usepackage{pbox}
\usepackage{algorithm}
\usepackage[noend]{algpseudocode}
\usepackage{algorithm}
\usepackage[noend]{algpseudocode}
\usepackage{array}

\usepackage{nicefrac}
\usepackage{mathtools}
\usepackage{multirow}

\usepackage{epstopdf}
\usepackage{xspace}
\usepackage{svg}

\usepackage{cite}

\makeatletter
\def\BState{\State\hskip-\ALG@thistlm}

\makeatother
\algnewcommand{\algorithmicgoto}{\textbf{go to}}%
\algnewcommand{\Goto}{\algorithmicgoto\xspace}%
\algnewcommand{\Label}{\State\unskip}

\usepackage[utf8]{inputenc}
\usepackage{tikz}
\usetikzlibrary{positioning,shadows}
\usetikzlibrary{shapes.geometric,arrows,calc,bending,arrows.meta}

\begin{document}
\title{Learning Self-Awareness for Autonomous Vehicles: Exploring Multisensory Incremental Models}

%
\author{Mahdyar~Ravanbakhsh, 
        Mohamad~Baydoun, 
        Damian~Campo, 
        Pablo~Marin, 
        David~Martin, 
        Lucio~Marcenaro, 
        and Carlo~Regazzoni 
\thanks{Mahdyar Ravanbakhsh, Mohamad Baydoun, Damian Campo,  Lucio Marcenaro, Carlo Regazzoni, are with the Department of Electrical, Electronics and Telecommunication Engineering and Naval Architecture, University of Genoa, Genoa 16145, Italy. (e-mail: 
mahdyar.ravan@ginevra.dibe.unige.it;
mohamad.baydoun@ginevra.dibe.unige.it;
damian.campo@ginevra.dibe.unige.it; lucio.marcenaro@unige.it; carlo.regazzoni@unige.it).}
\thanks{Pablo Marin and David Martin are with the Intelligent Systems Laboratory, Department of Systems Engineering and Automation, Universidad Carlos III de Madrid, 28911 Leganes, Spain (e-mail: pamarinp@ing.uc3m.es; dmgomez@ing.uc3m.es).}
}
\maketitle

\begin{abstract}
The technology for autonomous vehicles is close to replacing human drivers by artificial systems endowed with high-level decision-making capabilities. In this regard, systems must learn about the usual vehicle's behavior to predict imminent difficulties before they happen. An autonomous agent should be capable of continuously interacting with multi-modal dynamic environments while learning unseen novel concepts. Such environments are not often available to train the agent on it, so the agent should have an understanding of its own capacities and limitations. This understanding is usually called self-awareness. This paper proposes a multi-modal self-awareness modeling of signals coming from different sources. This paper shows how different machine learning techniques can be used under a generic framework to learn single modality models by using Dynamic Bayesian Networks. 
In the presented case, a probabilistic switching model and a bank of generative adversarial networks are employed to model a vehicle's positional and visual information respectively. 
Our results include experiments performed on a real vehicle, highlighting the potentiality of the proposed approach at detecting abnormalities in real scenarios. 
\end{abstract}


%
\IEEEpeerreviewmaketitle

\vspace{-0.25cm}
\printacronyms[include-classes=abbrev,name=Acronyms ]

\section{Introduction}
\label{sec:intro}
\begin{figure*}[h!]
\centering
	\includegraphics[width=\linewidth]{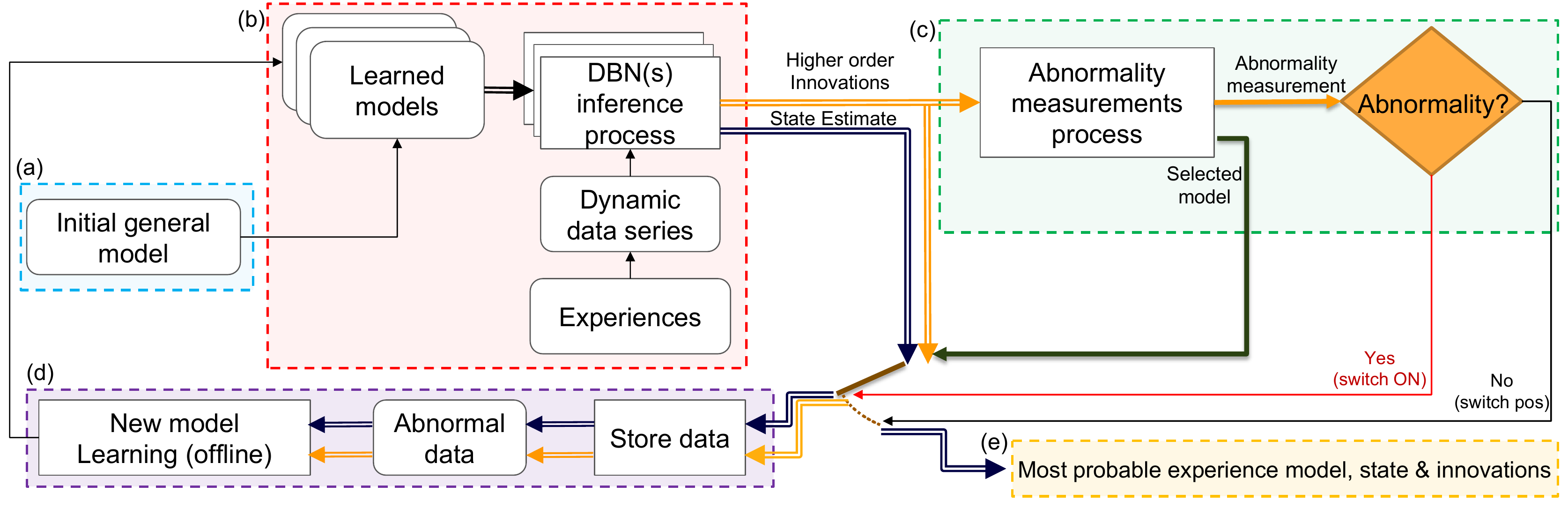}
	\caption{Block Diagram of Learning Process: the \ac{dbn} learning starts with an initial general model. New experiences that deviate from the general model (abnormalities) are detected. Identified abnormalities are stored and they can be used to learn a new model. Once the new learned model is available, the process can be iterated.}
	\label{fig:main_bd}
\end{figure*}

Autonomous vehicles are designed to take human error out of driving actions, which should help make self-driving vehicles safer, thus dramatically reducing the number of road accidents. Nowadays, methods based on machine learning techniques together with signal processing are playing an increasingly substantial role in this domain and contributing to general topics such as monitoring driver distraction \cite{Miyajima}, Smart Driver Monitoring \cite{Aghaei, 5462881}, vehicle infrastructure support and planning/monitoring \cite{Hult,8344848}, and detecting abnormalities automatically from multi-sensory data \cite{Campo2019,icassp2018,ravanbakhsh2017training}.


Systems with a single modality generally lack the robustness and reliability required in real-word applications \cite{8771378}. To tackle this issue, we propose a cross-modality structure using an agent's internal and external sensory data. Our approach is motivated by the process of problem-solving in humans, where information from various modalities, e.g., vision and language~\cite{martin2016language,6111485}, or different domains, e.g., brain and environment~\cite{6046230}, is combined to improve the decision-making.

\ac{sa} refers to the system's capability to recognize and predict its own states, possible actions, and the result of these actions on the system itself and its surroundings \cite{Schlatow2017}. By using \ac{sa} models, the agent gains the ability to both \emph{predict} the future evolution of a situation and \emph{detect} situations potentially unmanageable.  This ``sense of the limit'' is considered as a level of \ac{sa} that allows an agent to predict potential changes concerning its previous experiences to involve a human operator for support in due time. In this sense, the capability of detecting novel situations is an essential feature of \ac{sa} models as it allows autonomous systems to anticipate their situational/contextual states and improve the effectiveness of the decision-making sub-modules \cite{Campo2017,Olier2017a,icassp2018}. Our model is not a fully autonomous driving system; however, it is an aware system that can provide suggestions, predictions, and detect the possible deviation from previously learned tasks/situations. 


The work in \cite{icassp2018} introduces an \ac{sa} model consisting of two layers: \ac{sl} and \ac{pl}. The analysis of observed moving agents for learning normal/abnormal dynamics in a given scene represents an emerging research field \cite{Bastani2016, Morris2008, Moll2010, wacv_plug, nabi2013temporal, lin2017tube, sebe2018abnormal, sabokrou2016fully, mousavi2015abnormality, mousavi2015crowd, rabiee2016novel, rabiee2016crowd,ravanbakhsh2018learning,rabiee2018detection,ravanbakhsh2018fast}. 
An agent's planned activity can be modeled as a sequence of organized state changes (actions) performed to achieve a task under a particular context. This set of actions can be learned from observations that, in turn, can be clustered into sequential discrete patterns. The availability of a plan that associates the current state with an action model facilitates the detection of normal/abnormal situations in future repetitions of the same task. This paper complies with the definition of abnormality offered by, \cite{Ramik2014}, which defines it as a behavior (set of observations) that has not been seen before.

An active \ac{sa} action plan can be seen as a dynamical filter that predicts state behaviors using dynamic and observation models. \ac{sds}s \cite{Fox2011} are well-known probabilistic graphical models \cite{Koller2009} capable of managing discrete and continuous variables jointly in dynamical filters. Probabilistic models 
have demonstrated their usefulness at moment of modeling dependencies between different variables \cite{Vomlel2012}. In \cite{xu2017}, a probabilistic approach is taken for tracking random irregularities that evolve through time in vehicles/tracks.

This paper proposes a switching \ac{dbn2} that dynamically estimates future and present states at continuous and discrete levels. In \cite{Bastani2016}, \ac{sds}s have been used successfully to improve decision making and tracking capabilities. Moreover, the ability of probabilistic models to manage discrete representations enables the incorporation of semantic concepts in autonomous systems \cite{Huang2006,Li2008,Andrews2010}, which is useful feature for communicating/cooperating with humans. In \cite{Zhang2013,Hu2018}, authors propose probabilistic approaches that encode high-level semantics, e.g., drivers' intention, to model traffic vehicle data. 

The \ac{mjpf} \cite{Doucet2000} is one of the most used algorithms that takes advantage of learned hierarchical probabilistic models in the online \ac{sds} phase. \ac{mjpf} uses a combination of \ac{kf} and \ac{pf} to predict and update jointly continuous and discrete state space posterior probabilities. This paper employs an \ac{mjpf} to exploit learned knowledge from the \ac{sl} that includes the capability to self-detect abnormality situations.

An agent can infer dynamic models through internal variables, which represent private information only accessible by the agent itself. Detecting abnormalities is possible by training a \ac{sa} model on private (first person) multisensorial data. Such a model can be defined as the \ac{pl} of \ac{sa} \cite{icassp2018}. Most of the previous works rely on high-level supervision to learn \ac{sa} models for the \ac{pl} \cite{Olier2017a,Ramik2014,icassp2018}. However, this work proposes a weakly-supervised method based on a hierarchy of cross-modal \ac{gan}s \cite{NIPS2014_5423, icip17}. In this work, \ac{pl} data consists of first-person visual information taken by an operative agent. This visual information, coupled with the corresponding optical-flow data, is used to learn \ac{gan} models.

Our major contributions are \emph{(i)} proposing \ac{sds}s for learning several tasks incrementally in a semi-supervised fashion. \emph{(ii)} Introducing a multi-modal probabilistic model that can increase the capability of self-understanding and situational awareness in autonomous systems. Our model finds deviations (abnormalities) in human driving behaviors that may lead to detecting a new task, undesired situations, and even threats to the drivers' safety.

The proposed method includes two major phases: \emph{(i)} an incremental offline learning process and \emph{(ii)} an online testing procedure for detecting the possible abnormalities. The proposed offline incremental learning process for both \ac{sl} and \ac{pl} is described in Sec \ref{sec:method}, and the online abnormality detection phase is presented in Sec. \ref{sec:online_testing}. Sec. \ref{sec:res} introduces the employed dataset and shows the experimental results of training/testing phases. Sec. \ref{sec:Discussion} discusses the different components of our proposed method and Sec. \ref{sec:con} concludes the article.

\section{Multi-modal self-awareness modeling}
\label{sec:method}
\begin{table*}[h]
\centering
\begin{tabular}{lllll}\hline
{\textbf{Phases}} & \textbf{Steps/Components} &\textbf{SL(low dimensional data)} & \textbf{PL(high dimensional data)} & \textbf{Corresponding block(s)}\\ \cline{1-5}
  \multirow{5}{*}{Train (offline)} & input & positional and velocity data & first person visual data and optical-flow & b (training data series)\\ \cline{2-5}
  & initial filter & \ac{umkf} & reference \ac{gan} (linear movement) & a\\ \cline{2-5}
  & incremental & learning motivated KFs & detecting/learning outliers & d\\ \cline{2-5}
  & output& probabilistic graphical model & set of cross-modal \ac{gan}s & c, e \\ \cline{2-5}
  & final filter/model & MJPF & hierarchy of \ac{gan}s& b (set of learned models) \\ \cline{1-5}
  \multirow{3}{*}{Test (online)} & input & new experience positional data series &  new experience visual information & b (testing data series)\\ \cline{2-5}
  & output& positional state estimation &  visual prediction & c , e\\ \cline{2-5}
  & measurement & high level innovation & high level innovation & c \\ \cline{1-5} 
\end{tabular}
	\caption{Phases/components of the proposed method concerning the \ac{sa} modalities. See Fig. \ref{fig:learning_bd} for corresponding blocks.}
	\label{tab:summary}
\end{table*}

This work focuses on the incremental learning of dynamic models from data acquired through agent's experiences, facilitating the building of \ac{sa} models. This paper considers the actions of a human driver while performing different tasks observed from the vehicle's ``First Person" viewpoint. We propose an incremental adaptive process that allows the agent (vehicle) to learn switching DBN models from video and positional data. These models can predict (i.e., to generate) new observed situations and adaptively estimate the current states by filtering data based on the most fitting model (i.e., to discriminate). In other words, DBN models facilitate the prediction of situations different from the dynamic reference equilibrium (i.e., previously learned models).

Our \ac{sa} model consists of two levels: \ac{pl} and \ac{sl}. For both of them, a unified learning procedure is designed, see Fig. \ref{fig:main_bd}. The proposed incremental process (see Tab. \ref{tab:summary}) adds generative and discriminative knowledge to already known experiences. Statistically significant deviations from such known situations are recognized as abnormal situations whose characteristics are captured by the newly learned models.  

In our experiments, \ac{pl} and \ac{sl} are learned based on visual perception (first-person vision data only available to the agent) and localization (third-person vision), respectively. A probabilistic framework based on a set of switching dynamical models \cite{Doucet2000,fusion_valentina,6747301} is used to learn the \ac{sl}'s models incrementally. On the other hand, a bank of cross-modal \ac{gan}s is employed for learning \ac{pl} models. The rest of this section introduces a general procedure to learn dynamic models from a set of observed experiences and also presents a detailed explanation of learning processes for each level of \ac{sa}.

\begin{figure*}
    \centering
	\includegraphics[width=1\linewidth]{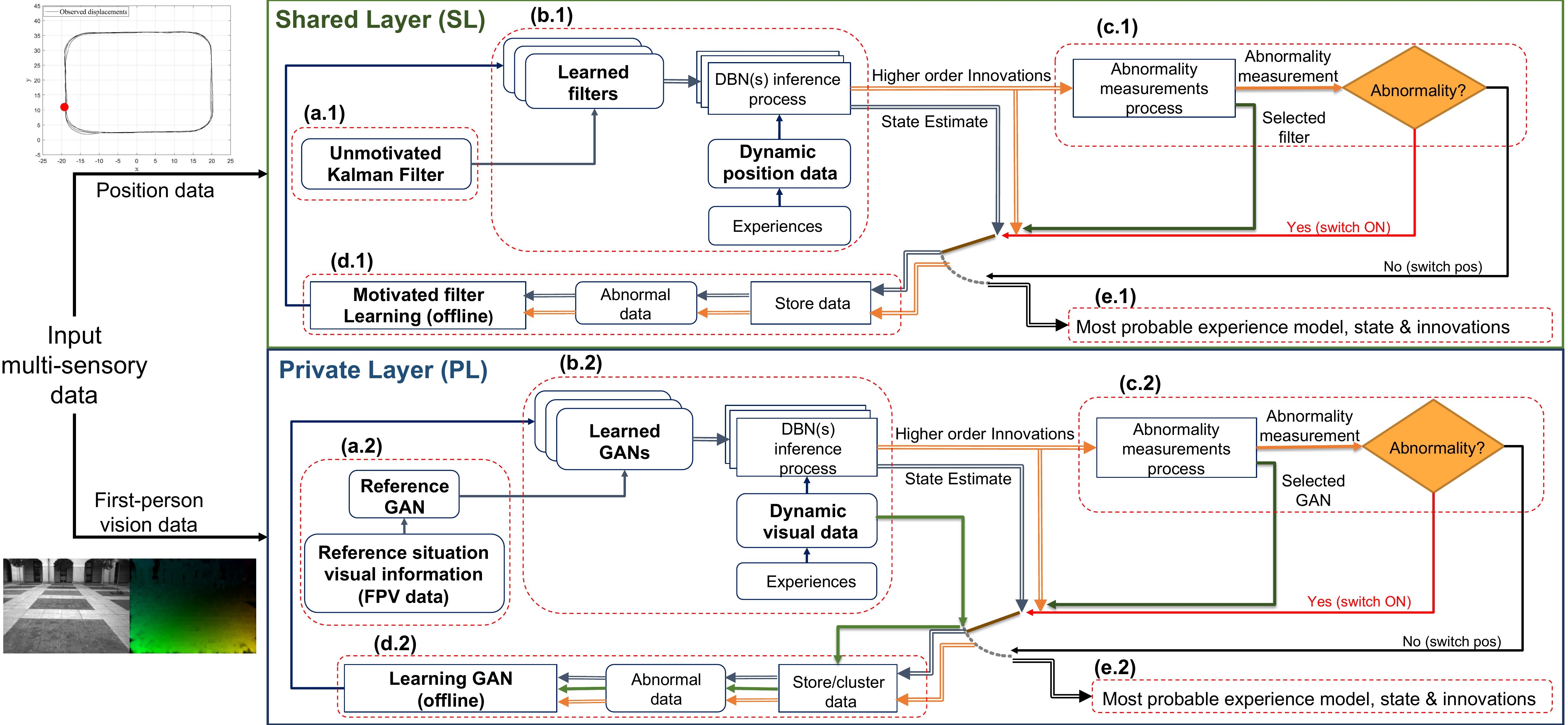}
	\caption{Learning Process in \ac{sl} and \ac{pl}: for each experience, \ac{sl} and \ac{pl} use respectively positional and video data and as inputs. The structure of both layers is based on a generic incremental adaptive training process that allows them to increase their self-awareness capabilities by learning new experiences observed from multi-sensory sources.}
	\label{fig:learning_bd}
\end{figure*}

\subsection{Learning SL and PL: an off-line process}
\label{sec:idea}
Fig. \ref{fig:main_bd} shows a flow chart describing our proposed adaptive learning process that enables the incremental learning of additional knowledge necessary to describe new observed situations. By encoding new knowledge into the agent's conditions of equilibrium trough probabilistic switching models, it is possible to increase the \ac{sa} level of the agent. 

The incremental learning of new models starts by observing dynamic data series related to a given experience (Fig. \ref{fig:main_bd}-b). Such a data is filtered by using an initial reference model (Fig. \ref{fig:main_bd}-a) and by keeping track of deviations w.r.t it. The initial reference model consists of a simple filter whose dynamic model describes a basic dynamic condition of equilibrium. In the case of \ac{sl}, where low dimensional data, i.e., 2D-locations and velocities, are considered, the initial filter corresponds to a \ac{kf} that assumes the agent remains static. This filter makes reasonable predictions when there are no forces affecting the state of the agent, i.e., in case the agent does not interact with its surroundings, suggesting that agent motions can be approximated by small random noise oscillations of the state. In the \ac{sl}, noises can be produced either by the agent or external entities, producing noisy data series of sparse measurements. Accordingly, we define it as the \ac{umkf}. Such a reference filter is illustrated in Fig. \ref{fig:learning_bd}-a.1. \ac{umkf} assumes $\dot{x} = \omega$, where $\dot{x}$ is the agent's velocity and $\omega$ is the perturbation error. 

When the \ac{umkf} is applied to a motivated experience, it produces a set of errors due to the fact that the agent follows certain motivations modeled as surroundings' forces acting on it. In the \ac{pl}, a pre-trained reference \ac{gan} is used as an initial general model (see Fig. \ref{fig:learning_bd}-a.2), which encodes an experience where the agent moves linearly on a clear path. Similar to the \ac{umkf} model, the reference \ac{gan} assumes a dynamic equilibrium exists between the environment and the agent but on a different modality, i.e., first-person video data). In this case, the condition of equilibrium corresponds to the agent's visual data supporting linear motion towards a point in the environment. 
Such a point can be seen as a stationary center of force that attracts the agent linearly. In case other forces interact with the agent, the dynamics of visual data would vary from the linear movement, producing a set of prediction mismatches (errors) between the \ac{gan} filter estimations and the new observations. By considering the set of errors (here defined as innovations, using terminology from \ac{kf}s) obtained from \ac{umkf} and the reference \ac{gan}, cumulative probabilistic tests can be designed for both \ac{sl} and \ac{pl} to evaluate whether a data series corresponds to an abnormal experience.

Collected innovations can be used to decide when to store data to learn a new filter. Each new filter represents a new equilibrium condition encoding a set of stationary forces, see Fig. \ref{fig:main_bd}-c. 
The abnormality measurement process block ranks innovations and computes abnormality signals, enabling the selection of the most probable model among those learned from previous experiences. The most fittest models represent an \ac{sds} in case of \ac{sl} and a couple of cross-modal \ac{gan}s for \ac{pl}. Abnormality signals can drive soft decision processes \cite{mazzu2016cognitive} and are used to learn new models, see Fig. \ref{fig:main_bd}-d. 

The abnormality detection procedure enables defining an incremental process similar to Dirichlet \cite{antoniak1974mixtures} (stick-breaking processes \cite{sethuraman1994constructive}, Chinese restaurant \cite{aldous1985exchangeability}). The proposed abnormality measurements determine the choice about whether new observations can be described by already available experiences (learned equilibrium conditions) embedded into DBN (Fig. \ref{fig:main_bd}-e) or there is a need to learn a new one (Fig. \ref{fig:main_bd}-d). Data from new experiences can be structured into multiple partitions of the state-innovation, where the correlations between states and innovations facilitate clustering the new data into classes that define a new learned model. Fig. \ref{fig:main_bd}-d shows the procedure of learning new models from state-innovation pairs. Such a data couple establishes a relationship between the states and error measurements obtained through the initial models. 
In the case of \ac{sl}, the new learning model process generates a set of regions that segment the state space motion depending on innovations (See Fig. \ref{fig:learning_bd}-d.1). For the \ac{pl}'s case, it is considered a clustering of consecutive images and their corresponding optical flows based on a similarity measurement (\emph{i.e.}, local innovation), see Fig. \ref{fig:learning_bd}-d.2. Detected regions can form an explicit (in \ac{sds}) or implicit (in \ac{gan}s) vocabulary of switching variables employed by learned model to describe new experiences. Note that, in the case of \ac{pl}, there is a need for accessing dynamic visual information to train a new set of \ac{gan}s. Hence, as shown in Fig. \ref{fig:learning_bd}-d.2, for detected new clusters based on state-innovation pairs, the original data is used directly for training the new \ac{gan}s. Accordingly, \ac{pl}'s models are all related to different effects of forces distinct from the one producing a linear motion in a video sequence, e.g., a curving motion will generate a new \ac{gan} model.

\subsection{Mathematical modeling of spatial activities for SL layer}
\label{subsec:learn_SL}
In \ac{sl}, The main idea consists of learning a switching DBN for tracking and predicting the dynamical system over time, see Fig. \ref{fig:dbn_pl} (b). Hence, arrows represent conditional probabilities between the involved variables. Vertical arrows facilitate to describe causalities between both continuous and discrete levels of inference and observed measurements.

\noindent{\textbf{Dynamic modeling.}} The \ac{sl} level of \ac{sa} focuses on the analysis of agents' positional information for understanding their dynamics in a given scene through a set of \ac{sl}'s models $\boldsymbol{M} = \{m\}_{m = 1,\dots, M}$. As it is well known, the dynamics of an agent can be described by hierarchical probabilistic models consisting of continuous and discrete random variables. Accordingly, an agent's dynamic model can be written as:
\begin{equation}\label{eq1.3}
X_{k+1} = AX_{k} + BU_{S^m_{k}} + w_k,
\end{equation}  
where $X_{k}$ represents the agent's state composed of its coordinate positions and velocities at a time instant $k$, such that $X_k = [\boldsymbol{x} \hspace{0.2cm} \boldsymbol{\dot{x}}]^\intercal$. $\boldsymbol{x} \in \mathbb{R}^d$ and $\boldsymbol{\dot{x}} \in \mathbb{R}^d$. $d$ represents the dimensionality of the environment. $A = [A_1 \hspace{0.2cm} A_2]$ is a dynamic model matrix: $A_1 = [I_d \hspace{0.2cm}  0_{d,d}]^\intercal$ and $A_2 = 0_{2d,d}$. $I_n$ represents a square identity matrix of size $n$ and $0_{l,g}$ is a $l \times g$ null matrix. $w_k$ represents the prediction noise which is here assumed to be zero-mean Gaussian for all variables in $X_k$ with a covariance matrix $Q$, such that $w_k \sim \mathcal{N}(0,Q)$.  

In Eq.\eqref{eq1.3}, $B = [I_2\Delta k \hspace{0.2cm} I_2]^\intercal$ is a control input model and $\Delta k$ is the sampling time. $U_{S^m_{k}} = [\dot{x}_k$, $\dot{y}_k]^T$ is a control vector that encodes the expected agent's velocity when its state belongs to a discrete region $S^m_k \in \boldsymbol{S}^m$. Discrete regions associated with a model $m$ can be represented as:
\begin{equation}\label{eq1.4}
\boldsymbol{S}^m = \{S^{m,l^m}\}_{l^m = 1,\dots,L^m},
\end{equation}
where $l^m$ and $L^m$ represent the index and the maximum number of super-states respectively. Additionally, a threshold value is defined where linear continuous models of super-states $\boldsymbol{S}^m$ are valid. Such a threshold is defined as:
\begin{equation}\label{eq1.6}
\psi_{\boldsymbol{S}^m} = E(d_{\boldsymbol{S}^m}) + 3\sqrt{V(d_{\boldsymbol{S}^m})},
\end{equation}  
where $d_{\boldsymbol{S}^m}$ is a vector containing all distances between neighboring super-states, $E(\cdot)$ receives a vector of data and calculates its mean and $V(\cdot)$ its variance. The threshold value in Eq.\eqref{eq1.6} defines a certainty boundary that determines where the model is valid, determining the extend of the \textit{knowledge} captured by the model. 

Let $Z_k$ be the observed agent's position at the instant $k$ while executing a given experience, see Fig. \ref{fig:learning_bd}-b.1. It is assumed a linear relationship between measurements and agent's states, such that:
\begin{equation}
\label{eq1.1}
Z_{k} = HX_{k} + \nu_k,
\end{equation}
where $H = [I_d \hspace{0.2cm} 0_{d,d}]$ is an observation matrix that maps states onto observations and $\nu_k$ represents the measurement noise produced by the sensor device which is here assumed to be zero-mean Gaussian with a covariance matrix $R$, such that, $v_k^m \sim \mathcal{N}(0,R)$. By considering the dynamical filter in Eq.\eqref{eq1.3}, it is possible to estimate the agents' velocity by using the KFs' innovations produced by a model $m$, such that:
\begin{equation}
\label{eq1.2}
v_k^m= \frac{Z_k - H\hat{X}^m_{k+1|k}}{\Delta k},
\end{equation}
where $\hat{X}^m_{k+1|k}$ is the state space estimation under the model $m$ at the time $k+1$ given observations until time $k$. $\Delta k$ is the sampling time.
By considering a 2-dimensional space, i.e., $d=2$, the agent's state can be written as: $X_k = [x_k,y_k,\dot{x}_k,\dot{y}_k]^\intercal$.

\noindent{\textbf{Initial model.}}
 The initial model $m=0$ is a situation where the agent keeps the same position over time. A \ac{kf} based on an ``unmotivated model'' in Fig. \ref{fig:learning_bd}-a.1 is used to tracking agents. 

The model $m = 0$ contains only one super-state $\boldsymbol{S^0} = \{S^0_1\}$, leading to $U_{S^0_1} \sim 0$. By relaxing $BU_{S^m_{k}}$ in Eq.\eqref{eq1.3}, we obtain:  $X_{k+1} = AX_{k} + w_{k}$, where the agent is assumed to move only under random noisy fluctuations $w_k$. By applying Eq.\eqref{eq1.2}, innovations obtained from the initial model $m=0$ can be collected and used to create new models incrementally. 

\noindent{\textbf{Creating models incrementally.}}
As shown in Fig. \ref{fig:main_bd}-c, during the inference, there are two different possible situations.\\
\emph{i) Normality:} the observation can be fitted and predicted with the current learned model(s). In this case, there is no need to learn new models and Eq. \eqref{eq1.3} is used to infer future states.\\
\emph{ii) Abnormality:} the current observation does not fit in the existing model(s), which means it is out of the boundary defined in Eq.\eqref{eq1.6}, where a random filter $U_{S^m_{k}} = 0_{2,1}$ is applied. In this case, the abnormal data is used to learning a new model $m+1$. For \ac{sl}, this corresponds to learning a \ac{mkf} in Fig. \ref{fig:learning_bd}-d.1, where the agent's dynamics can be described by quasi-constant velocity models, i.e., $U_{S^m_{k}} \neq 0$ in Eq.\eqref{eq1.3}.

The state information of instances detected as abnormal is collected to learning new models. We employ a \ac{som} \cite{Kohonen2001} that receives stored abnormal states $X_k$ and generates a set of neurons encoding similar information (quasi-constant velocities). Consequently, the set of super-states will be updated, such that:
\begin{equation}
\label{eq:inc}
\boldsymbol{S}^{m+1}  = \{S^{m+1,l^{m+1}}\}_{l^{m+1} = 1,\dots,L^{m+1}}.  
\end{equation}

The clustering process prioritizes similar velocities (actions) based on a weighted distance. Consequently, Eq. \eqref{eq1.5} shows a distance function that uses the weights $\beta$ and $\alpha$ employed to train the \ac{som}, where $\beta + \alpha = 1$ and  $\alpha > \beta$ to favor clustering of patterns with small differences in velocity, such that:

\begin{equation}\label{eq1.5}
{d}(\tilde{X},\tilde{Y}) =\sqrt{(\tilde{X}-\tilde{Y})^\intercal D(\tilde{X}-\tilde{Y})},
\end{equation} 
where $D = [\mathcal{B} \hspace{0.2cm} \mathcal{A}]$. $\mathcal{B} = [\beta I_2 \hspace{0.2cm} 0_{2,2}]^\intercal$, $\mathcal{A}  = [0_{2,2} \hspace{0.2cm} \alpha I_2]^\intercal$. $\tilde{X}$ and $\tilde{Y}$ are both 4-dimensional vectors of the form $[x \hspace{0.2cm} y \hspace{0.2cm} \dot{x} \hspace{0.2cm} \dot{y}]^\intercal$.

By analyzing the activated super-states over time while executing a certain activity, it is possible to obtain a set of temporal transition matrices $T^m_{t}$. Those matrices encode the probabilities of passing or staying between super-states depending on the time $t$ that the agent has spent in a super-state while model $m$ is applied. Transition matrices facilitate the inference of next super-states given the current one, i.e., $p(S^m_k|S^m_{k-1},t)$.

Each new model is identified in an unsupervised fashion and added to the model set $\boldsymbol{M} = \{m\}_{m = 1,\dots, M}$. While the agent perform a task, all transition matrices $T^m_{t}$ apply in parallel and the model with minimum error $m_{k,(min)}$ is selected as the correct one, such that: 
\begin{equation}\label{eq:minError}
m_{k,(min)} = \underset{m}\argmin\big(v_k^m\big).
\end{equation}

In case $v_k^{m_{k,(min)}} > \psi_{\boldsymbol{S}^m}$, the selected model $m_{k,(min)}$ is not valid and a new model must be added to the available ones, see Eq.\eqref{eq:inc}. Similarly, a new set of temporal transition matrices $T^{m+1}_{t}$ are calculated for the new detected model.


\colorlet{rectangle edge}{blue!50}
\colorlet{rectangle area}{red!20}

\tikzset{filled/.style={fill=rectangle area, draw=rectangle edge, thick},
    outline/.style={draw=rectangle edge, thick}}

\begin{figure}
\begin{minipage}{.45\linewidth}
\scalebox{0.55}{
\begin{tikzpicture}[->,
roundnode/.style={circle, draw=black!60, fill=green!5, very thick, minimum size=1.3cm},
imaginarynode/.style={circle, very thick, minimum size=1mm},
rectanglenode/.style={rectangle, draw=black!60, fill=blue!5, very thick, minimum size=1cm}
]
\filldraw[color=red!60, fill=red!5, very thick, rounded corners=15pt](-2,-1) rectangle (5.5,3.2) node[rotate=-90]at (6,1.2) {\textbf{\Large Hidden Markov Model}};
\draw[color=blue!60, very thick, rounded corners=15pt](-2,0.8) rectangle (5.5,-5.8) node[rotate=-90]at (6,-3.2) {\textbf{\Large GANs}};
\node[roundnode]        (xk_1)                                  {${\cal X}_{k-1}$};
\node[roundnode]        (sk_1)                [above=of xk_1]   {$C_{k-1}$};
\node[roundnode]        (xpk_1)               [below=of xk_1]   {${\cal X}^{\cal P}_{k-1}$};
\node[roundnode]        (zk_1)                [below=of xpk_1] {${\cal Z}_{k-1}$};

\node[imaginarynode]    (sk_2)       [left=of sk_1]     {};
\node[imaginarynode]    (xk_2)       [left=of xk_1]     {};
\node[imaginarynode]    (xpk_2)      [left=of xpk_1]    {};
\node[imaginarynode]    (zk_2)       [left=of zk_1]     {};

\node[roundnode]    (xk)                [right=0cm and 2cm of xk_1]     {${\cal X}_{k}$};
\node[roundnode]    (xpk)               [right=0cm and 2cm of xpk_1]    {${\cal X}^{\cal P}_{k}$};
\node[roundnode]    (sk)                [above=of xk]                   {$C_{k}$};
\node[roundnode]    (zk)                [below=of xpk]                   {${\cal Z}_{k}$};

\node[imaginarynode]    (sk_n)       [right=of sk]  {};
\node[imaginarynode]    (xk_n)       [right=of xk]  {};
\node[imaginarynode]    (xpk_n)      [right=of xpk] {};
\node[imaginarynode]    (zk_n)       [right=of zk]  {};

\draw[dashed] (sk_2.east) -- (sk_1.west);
\draw[dashed] (xk_2.east) -- (xk_1.west);
\draw[dashed] (xpk_2.east) -- (xpk_1.west);
\draw[dashed] (zk_2.east) -- (zk_1.west);

\draw[->] (sk_1.south) -- (xk_1.north);
\draw[->] (xk_1.south) -- (xpk_1.north);
\draw[->] (xpk_1.south) -- (zk_1.north);

\draw[->] (sk_1.east) -- (sk.west)  node[midway, above] {$p(C_k|C_{k-1})$};

\draw[->] (xk_1.east) -- (xk.west) node[midway, above] (of_k) {{$p({\cal X}_k={\cal X}_{k-1})$}};

\draw[->] (sk.south) -- (xk.north) node[midway, right] {$p({\cal X}_k|C_k)$};
\draw[->] (xk.south) -- (xpk.north) node[midway, right] {$p({\cal X}_k|{\cal X}^{\cal P}_{k})$};
\draw[->] (xpk.south) -- (zk.north) node[midway, right] {$p({\cal X}^{\cal P}_{k}|{\cal Z}_{k})$};
\draw [->] (xk.south west) to [out=250,in=110] (zk.west);
\node[imaginarynode]    (1)       [below=of of_k]     {$p({\cal X}_{k}|{\cal Z}_{k})$};

\draw[dashed] (sk.east) -- (sk_n.west);
\draw[dashed] (xk.east) -- (xk_n.west);
\draw[dashed] (xpk.east) -- (xpk_n.west);
\draw[dashed] (zk.east) -- (zk_n.west);


\end{tikzpicture}}
\centering
(a)
\end{minipage}
 \hspace*{0.5cm}
\begin{minipage}{.45\linewidth}
	\scalebox{0.55}{
\begin{tikzpicture}[
roundnode/.style={circle, draw=black!60, fill=green!5, very thick, minimum size=1.3cm},
imaginarynode/.style={circle, very thick, minimum size=1mm},
]
\filldraw[color=red!60, fill=red!5, very thick, rounded corners=15pt](-2,-0.8) rectangle (5.5,3.2) node[rotate=-90]at (6,1.8) {\textbf{\Large Particle Filter}};
\draw[color=blue!60, very thick, rounded corners=15pt](-2,0.8) rectangle (5.5,-3.2) node[rotate=-90]at (6,-1.5) {\textbf{\Large Kalman Filter}};
\node[roundnode]        (xk_1)                              {$X_{k-1}$};
\node[roundnode]        (sk_1)                [above=of xk_1] {$S_{k-1}$};
\node[roundnode]        (zk_1)                [below=of xk_1] {$Z_{k-1}$};

\node[imaginarynode]    (sk_2)       [left=of sk_1] {};
\node[imaginarynode]    (xk_2)       [left=of xk_1] {};
\node[imaginarynode]    (zk_2)       [left=of zk_1] {};

\node[roundnode]    (xk)                [right=0cm and 2cm of xk_1] {$X_{k}$};
\node[roundnode]    (sk)                [above=of xk] {$S_{k}$};
\node[roundnode]    (zk)                [below=of xk] {$Z_{k}$};

\node[imaginarynode]    (sk_n)       [right=of sk] {};
\node[imaginarynode]    (xk_n)       [right=of xk] {};
\node[imaginarynode]    (zk_n)       [right=of zk] {};

\draw[dashed] (sk_2.east) -- (sk_1.west);
\draw[dashed] (xk_2.east) -- (xk_1.west);
\draw[dashed] (zk_2.east) -- (zk_1.west);

\draw[->] (sk_1.south) -- (xk_1.north);
\draw[->] (xk_1.south) -- (zk_1.north);

\draw[->] (xk_1.east) -- (xk.west) node[midway, above] {$p(X_k|X_{k-1})$};
\draw[->] (sk_1.east) -- (sk.west)  node[midway, above] {$p(S_k|S_{k-1})$};

\draw[->] (sk.south) -- (xk.north) node[midway, left] {$p(X_k|S_k)$};
\draw[->] (xk.south) -- (zk.north) node[midway, left] {$p(Z_k|X_k)$};

\draw[dashed] (sk.east) -- (sk_n.west);
\draw[dashed] (xk.east) -- (xk_n.west);
\draw[dashed] (zk.east) -- (zk_n.west);


\end{tikzpicture}}
\centering
(b)
\end{minipage}
\caption{Proposed DBN switching models for (a) private layer, (b) shared layer}
\label{fig:dbn_pl}
\end{figure}
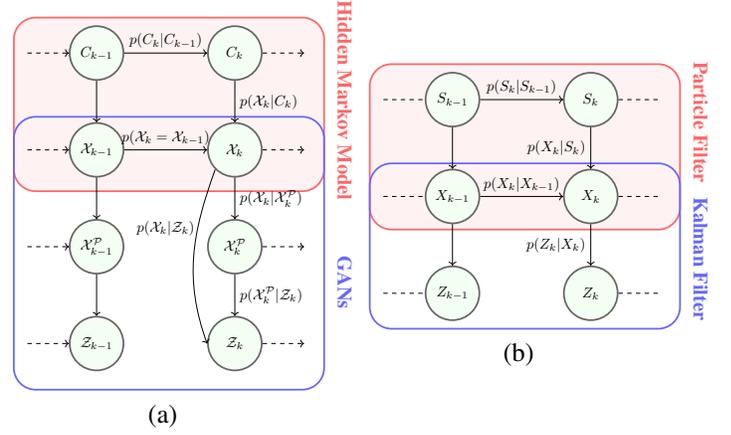
\subsection{Mathematical modeling of PL layer} 
\label{subsec:learn_PL}
\noindent{\textbf{Cross-modal GAN representation.}} Unlike the \ac{sl}, the \ac{pl} deals with high-dimensional visual information observed by the agent. Namely, a sequence of images (frames) ${\cal I}$, and their corresponding optical-flow maps (motion) ${\cal O}$. To model the \ac{pl} of \ac{sa}, a set of cross-modal GANs \cite{NIPS2014_5423} is trained to learn the normality from a set of visual data. Generative models try to maximize likelihood by minimizing the Kullback-Leibler distance between a given data distribution and the generator's distribution \cite{arjovsky2017towards}. GANs learn such a minimization during the training phase by an adversarial game between two networks: a generator ($G$) and a discriminator ($D$). 

As mentioned before, a \ac{sa} model has to be generative (predict multi-level, temporal data series based on previously learned knowledge) and discriminative (provide measurements to evaluate or select best-fitting models in new observations). In the case of GANs, Generative networks learn to predict. In particular, our predictions include the generation of the next image (frame) and optical-flow (motion map). This could be seen as a hidden state ${\cal X}^{\cal P}_{k}$ (see Fig. \ref{fig:dbn_pl}). The update task includes having the likelihood and prediction. In GANs, the likelihood is learned and approximated by the Discriminators. The outputs of the Discriminators are the encoded version of optical-flow ${\cal D}^{\cal O}$ and image ${\cal D}^{\cal I}$. In our approach, Discriminators' scores are used to approximate the distance between the likelihood and the prediction.

Intuitively, the encoded version of an image can be seen as the state in \ac{sl}, while the encoded version of optical-flow represents its motion. The error ${\cal E}$ can be seen as the distances between the encoded versions of prediction and the observation. Following the same intuition applied in \ac{sl}, we cluster the encoded version of images, motion and errors into a set of super-states. In light of the above, the states here can be defined as a function of image, motion, and error $f([{\cal D}^{\cal I},{\cal D}^{\cal O},{\cal E}])$ for any model (super-state).


\noindent{\textbf{Dynamic modeling.}}
GANs are commonly used to generate data (e.g., images) and trained using only unsupervised data. Supervisory information in a \ac{gan} is indirectly provided by an adversarial game between two independent networks: a generator ($G$) and a discriminator ($D$). During training, $G$ generates new data and $D$ tries to distinguish whether its input is real (i.e., a training image) or it was generated by $G$. The competition between $G$ and $D$ helps to boost the ability of both $G$ and $D$. For learning conditions of equilibrium, two channels are used as observations: appearance (raw-pixels) and motion (optical-flow images) for two different cross-channel tasks. In the first task, optical-flow images are generated from the original frames. In the second task, appearance information is estimated from an optical flow image.
Specifically, let ${\cal I}_k$ be the $k$-th frame of a training video and ${\cal O}_k$ the optical-flow obtained through ${\cal I}_k$ and ${\cal I}_{k+1}$. ${\cal O}_k$ is computed using \cite{brox2004high}.
For any given model $m' \in \boldsymbol{M'}$, where $\boldsymbol{M'} = \{m'\}_{m' = 1,\dots, M'}$, two networks are trained: ${\cal N}^{m':{\cal I} \rightarrow {\cal O}}$, which generates optical-flow from frames (task 1) and ${\cal N}^{m':{\cal O} \rightarrow {\cal I}}$, which outputs frames from optical-flow (task 2).
In both cases, inspired by \cite{DBLP:journals/corr/IsolaZZE16,icip17}, our architecture is composed of two fully-convolutional networks: the conditional generator $G$ and discriminator $D$. 
$G$ is the U-Net architecture \cite{DBLP:journals/corr/IsolaZZE16}, which is an encoder-decoder following with {\em skip connections} helping to preserve relevant local information. 
$D$ is the {\em PatchGAN} discriminator \cite{DBLP:journals/corr/IsolaZZE16}, which is based on a ``small'' fully-convolutional discriminator. 
$G$ and $D$ are trained using both a conditional \ac{gan} loss ${\cal L}_{cGAN}$ and a reconstruction loss ${\cal L}_{L1}$. In case of ${\cal N}^{m':{\cal I} \rightarrow {\cal O}}$, the training set is composed of pairs of frame-optical flow images
${\cal X} = \{ (F_t, O_t) \}_{t=1,...,N}$. 
${\cal L}_{L1}$ is given by: ${\cal L}_{L1}(x,y) =  ||y - G(x,z) ||_1$,
\begin{equation}
{\cal L}_{L1}(x,y) =  ||y - G(x,z) ||_1,
\end{equation}
\noindent
where $x = F_t$ and $y = O_t$,
while the conditional adversarial loss ${\cal L}_{cGAN}$ is:
\begin{equation}
\begin{split}
{\cal L}_{cGAN}(G,D)= 
\mathbb{E}_{(x,y) \in {\cal X}} [\log D(x,y)] +\\
\mathbb{E}_{x \in \{ F_t \},z \in {\cal Z}} [\log ( 1 - D(x,G(x,z)) )].
\end{split}
\end{equation}
In case of ${\cal N}^{m':{\cal O} \rightarrow {\cal I}}$, we define ${\cal X} = \{ (O_t, F_t) \}_{t=1,...,N}$. 
Additional details about the training can be found in \cite{DBLP:journals/corr/IsolaZZE16}. 
In the GANs' training phase, networks ${\cal N}^{m':{\cal I} \rightarrow {\cal O}}$ and ${\cal N}^{m':{\cal O} \rightarrow {\cal I}}$ learn a dynamic model for the continuous space.

The discrete level uses an encoded vector $C^{m'}_{k} = [D^{m':{\cal O} \rightarrow {\cal I}} ({\cal I}_k, {\cal O}_k), D^{m':{\cal O} \rightarrow {\cal I}} ( {\cal I}_k,  {\cal O}_k)]$, where $D^{m':{\cal O} \rightarrow {\cal I}}$ and $D^{m':{\cal I} \rightarrow {\cal O}}$ are the discriminator networks of ${\cal N}^{m':{\cal O} \rightarrow {\cal I}}$ and ${\cal N}^{m':{\cal I} \rightarrow {\cal O}}$, respectively. Encoded vectors represent the expected entity's motion when its state belongs to a region $C^{m'}_k$, where $m'$ is a given model. Discrete regions of a given model $m'$ are written as:
\begin{equation}\label{eq1.4}
\boldsymbol{C^{m'}} = \{C^{m',l^{m'}}\}_{l^{m'}=1,\dots,L^{m'}},
\end{equation}
where $C^{m'}_k \in \boldsymbol{C^{m'}}$ and $L^{m'}$ is the total number of super-states in a task $m'$. Additionally, a threshold is employed to define where linear continuous models of super-states $\boldsymbol{C}^{m'}$ are valid. The threshold can be written as:
\begin{equation}\label{eq1.6.2}
\psi_{\boldsymbol{C}^{m'}} = E({\cal D}_{\boldsymbol{C}^{m'}}) + 3\sqrt{V({\cal D}_{\boldsymbol{C}^{m'}})},
\end{equation}  
where ${\cal D}_{\boldsymbol{C}^{m'}}$ contains all cross-modal discriminators likelihoods over the super-states, $E(\cdot)$ and $V(\cdot)$ are defined in Eq.\eqref{eq1.6}. Note that Eq.\eqref{eq1.6} determines the extend of the \textit{knowledge} captured by GANs.  

\noindent{\textbf{Initial model.}}
GANs are trained in a weakly-supervised manner, their only supervision consists of a subset of normal data to train the first level of the hierarchy that we name \emph{reference GANs}, which corresponds to U-KF in case of \ac{sl} for low-dimensional data. The \emph{reference GANs} is trained to model an initial equilibrium condition where the agent moves linearly towards a fixed motivation point. The \emph{reference GANs} provide a baseline for the next levels of the models, in which all of them are trained in a self-supervised way. 

Similar to the \ac{sl} initial model, the \ac{pl} initial model $m' = 0$ contains a single super-state $\boldsymbol{C^0} = \{C^0_1\}$, containing the \emph{reference GANs}. Details of the training are shown in Alg. \ref{alg:hgan}. The inputs of the procedure are represented by two sets: ${\cal Z}$ represents an observation vector including all normal observations from the training data and  ${\cal V}_{m'}$, which is a subset of ${\cal Z}$. In case of the \emph{reference GANs}, the initial set ${\cal V}_0$ is used to train two cross-modal networks ${\cal N}^{0:{\cal I} \rightarrow {\cal O}}$, and ${\cal N}^{0:{\cal O} \rightarrow {\cal I}}$. Note that the only supervision consists of training the first model (\emph{reference GANs}) on the initial set ${\cal V}_0$. the next models are built from the supervision provided by the \emph{reference GANs}.

Moreover, the mismatches between the \emph{reference GANs} estimation and new observations lead to errors that, in turn, are used to detect new conditions of equilibrium. 

\noindent{\textbf{Creating models incrementally.}}
As described in Sec.\ref{sec:intro}, our method assumes that the distribution of the normality patterns has a high degree of diversity. To learn such a distribution, we suggest a hierarchical strategy for high-diversity areas by encoding the different distributions into the different levels, in which each subset of train data is used to train a different \ac{gan}. A recursive procedure is adopted to construct the proposed hierarchy of GANs. As shown in Alg. \ref{alg:hgan}, the input set ${\cal Z}$ includes a set of coupled Frame-Motion maps, where ${\cal Z} = \{ [{\cal I}_k, {\cal O}_k] \}_{k=1,...,N}$, and $N$ is the number of total train samples. Besides, the input ${\cal V}_{m'}$ is a subset of ${\cal Z}$, provided to train GANs for each model. 

After training ${\cal N}^{0:{\cal I} \rightarrow {\cal O}}$, and ${\cal N}^{0:{\cal O} \rightarrow {\cal I}}$, we input $G^{0:{\cal I} \rightarrow {\cal O}}$ and $G^{0:{\cal O} \rightarrow {\cal I}}$ using each frame ${\cal I}$ of the entire set ${\cal Z}$ and its corresponding optical-flow image ${\cal O}$, respectively. The generators predict Frame-Motion couples as:
\begin{equation}\label{eq:pred_gan}
\begin{multlined}
\quad {\cal X}^{0:\cal P} = \{ [{\cal P}^{0:\cal I}_{k} , {\cal P}^{0:\cal O}_{k}]\}_{k=1,...,N}\\ 
{\cal P}^{0:\cal I}_{k} = G^{0:{\cal O} \rightarrow {\cal I}} ( {\cal O}_k),\quad
{\cal P}^{0:\cal O}_{k} = G^{0:{\cal I} \rightarrow {\cal O}} ( {\cal I}_k),
\end{multlined}
\end{equation}
where ${\cal P}^{0:\cal I}_{k}$ and ${\cal P}^{0:\cal O}_{k}$ are the $k$-th predicted image and optical-flow, respectively. The encoded versions of observations ${\cal Z}$ are computed by the discriminator networks $D^0$:
\begin{equation}\label{eq:dist_gan_z}
\begin{multlined}
\quad{\cal D}^{0:\cal I} = \{D^{0:{\cal O} \rightarrow {\cal I}} ( {\cal I}_k,  {\cal O}_k)\}_{k=1,...,N} ,\\ 
{\cal D}^{0:\cal O} = \{D^{0:{\cal I} \rightarrow {\cal O}} ( {\cal O}_k,  {\cal I}_k)\}_{k=1,...,N},
\end{multlined}
\end{equation}
where ${\cal D}^{0:\cal I}$ and ${\cal D}^{0:\cal O}$ are the encoded version (from initial model $m'=0$) of the observed image and optical-flow, respectively.
Encoded distance maps ${\cal E}^{0}$ between observations $\cal Z$ and predictions $\cal P$ for both channels are computed as:
\begin{equation}\label{eq:dist_gan}
\begin{multlined}
\quad \quad  \boldsymbol{{\cal E}}^{0} = \{ [{\cal E}^{{0:}\cal I}_{k} , {\cal E}^{{0:}\cal O}_{k}]\}_{k=1,...,N},\\ 
\quad{\cal E}^{{0:}\cal I}_{k} = D^{0:{\cal O} \rightarrow {\cal I}} ( {\cal I}_k,  {\cal O}_k) -  
D^{0:{\cal O} \rightarrow {\cal I}} ( {\cal P}^{\cal I},  {\cal O}_k) ,\\\quad 
{\cal E}^{{0:}\cal O}_{k} = D^{0:{\cal I} \rightarrow {\cal O}} ( {\cal O}_k,  {\cal I}_k) -  
D^{0:{\cal I} \rightarrow {\cal O}} ( {\cal P}^{\cal O}_k,  {\cal I}_k).\quad 
\end{multlined}
\end{equation}
The distance maps $\boldsymbol{{\cal E}}^0$ represent a set of errors in the coupled image-motion states representation, The joint states $\{ [{\cal D}^{0:\cal I}_{k} , {\cal D}^{0:\cal O}_{k}, {\cal E}_{k}]\}_{k=1,...,N}$ input to a \ac{som} that clusters similar appearance-motion information. Similar to clustering position-velocity information in the shared layer, the proposed clustering groups the appearance-motion representations into a set of super-states. Specifically, the SOM's output is a set of neurons encoding the state information into a set of prototypes. Detected prototypes (clusters) provide the means of discretization for representing a set of super-states, and consequently, the set of super-states will be updated, such that:
\begin{equation}
\label{eq:inc}
\boldsymbol{C}^{m'+1}  = \{C^{m'+1,l^{m'+1}}\}_{l^{m'+1} = 1,\dots,L^{m'+1}},  
\end{equation}
where $L^{m'+1}$ is the number of detected clusters (super-states) for the new model(s).

It is expected that the clusters containing the training data present a low distance score due to low innovations between predictions and observations. This is the criteria to detect the new distributions for learning new GANs, in which the clusters with high average scores are considered as new distributions. The new detected distributions forming the new subsets ${\cal V}_{l}$ to train new networks ${\cal N}^{C^{m'}:{\cal I} \rightarrow {\cal O}}$, and ${\cal N}^{C^{m'}{\cal O} \rightarrow {\cal I}}$ for the new \ac{gan} models. New identified models add to the model set $\boldsymbol{M'} = \{m'\}_{m' = 1,\dots, M'}$. During a task performing by the agent, all the transition matrices $T^{m'}_{t}$ apply in parallel. The selected model (best-fitted model) will be selected based on the minimum error ${m'}_{k,(min)}$, such that: 
\begin{equation}\label{eq:minError}
{m'}_{k,(min)} = \underset{m'}\argmin\big({\cal E}_k^{m'}\big).
\end{equation}

In case ${\cal E}_k^{{m'}_{k,(min)}} > \psi_{\boldsymbol{C}^{m'}}$, the selected model ${m'}_{k,(min)}$ is not valid and a new model must be added to the available ones, see Eq.\eqref{eq:inc}. Similarly, a new set of temporal transition matrices $T^{m+1}_{t}$ are calculated for the new detected model.

This procedure continues until no new distribution is detected. \ac{gan}s and detected super-states in each level are stacked incrementally for constructing the entire \ac{gan} model set ${\cal H}_{C^{m'}}$. The incremental nature of the proposed method makes it capable of learning complex distributions of data in a self-supervised manner.

\begin{algorithm}
\caption{Incremental training: Hierarchy of GANs}
\label{alg:hgan}
\begin{algorithmic}[1]
\Require
\State $\psi_{\boldsymbol{C}^{m'}}:  \text{  Threshold parameter for train a new GAN}$
\State ${\boldsymbol{C}^{m'}} = \{C^0\}$
\State ${\cal Z}:$ \text{ Entire training sequences} ${{\cal Z} = \{ ({\cal I}_k, {\cal O}_k) \}_{k=1,...,N}}$ 
\State ${\cal V}_0: \text{  Subset of } {\cal Z}$
\State $m'=0: \text{  Counter of models}$
\Ensure
\State $\{{\cal H}_{\boldsymbol{C}^{m'}}\} \text{ Hierarchy of GANs}$
\Procedure{TRAINING OF CROSS-MODAL GANs}{}\label{marker}
\Label $\texttt{train:}$
\State $\text{Train networks } {\cal N}^{{m'}:{\cal I}\rightarrow {\cal O}}, {\cal N}^{{m'}:{\cal O} \rightarrow {\cal I}}\text{, with } {\cal V}_{m'}$
\State $\{{\cal H}_{\boldsymbol{C}^{m'}}\} \gets \text{Trained networks }{\cal N}^{{m'}:{\cal I}\rightarrow {\cal O}}, {\cal N}^{{m'}:{\cal O} \rightarrow {\cal I}} $
\State ${\cal X}^{{m'}:\cal P} \gets {G}^{m'} ({\cal Z}) \text{: predictions}$
\State ${\cal D}^{m'} \gets {D}^{m'} ({\cal Z}) \text{: encoded observation}$
\State ${\cal E}^{m'} \gets ||{D}^{m'} ({\cal Z}) - {D}^{m'} ({\cal X}^{{m'}:\cal P})||_1 \text{: error}$
\State ${\cal X} \gets [{\cal D}_{m'}, {\cal E}_{m'}] \text{: states}$
\State $\text{Clustering states: }SOM({\cal X}) \text{: super-states } \boldsymbol{C}^{m'}$

\For{$\text{each identified cluster}$}
\State $\mu \gets \text{Average score maps in this cluster}$ 

\If{$\mu \ge \psi_{\boldsymbol{C}^{m'}} $}
\State $ m'= m' +1 $
\State ${\cal V}_{m'} \gets \text{Samples from cluster } C^{m'}\text{ in }{\cal Z}$
\State \Goto \texttt{train}
\EndIf
\EndFor
\State \textbf{return} $\{{\cal H}_{\boldsymbol{C}^{m'}}\}$
\EndProcedure
\end{algorithmic}
\end{algorithm}



\section{Abnormality detection: an online application}
\label{sec:online_testing}

\subsection{Shared layer: on-line testing MJPF}
\label{subsubsec:MJM}
\begin{figure}[t]
	\centering
	\includegraphics[width=\linewidth]{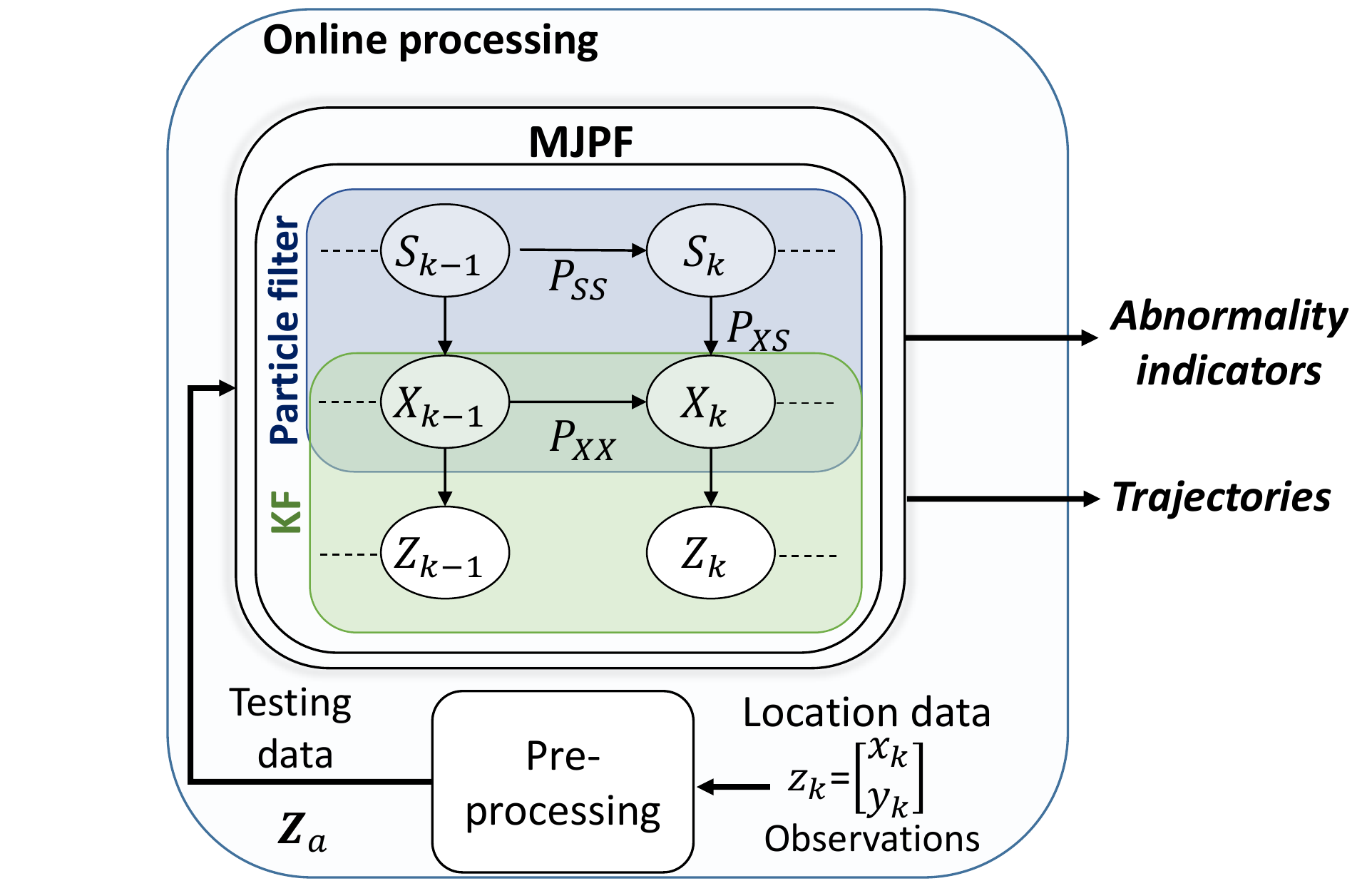}
	\caption{Online phase: An \ac{mjpf} makes inferences on the DBN}
	\label{fig:10}
\end{figure}

As mentioned before, the \ac{sl} model can be learned using a \ac{sds}, see Fig.  \ref{fig:dbn_pl} (b). Such a model includes a discrete set of state regions subspaces corresponding to the switching variables. The quasi-constant velocity model described before in Eq. \eqref{eq1.3}, is employed in each region to describe the relation between consecutive states in time. A further learning step facilitates to obtain the temporal transition matrices between super-states. An \ac{mjpf} is used to infer posterior probabilities on discrete and continuous states iteratively, see Fig.  \ref{fig:10}. Our \ac{mjpf} uses a \ac{pf} at the discrete level, embedding in each particle \ac{kf}. In other words, each particle in the \ac{pf} uses a \ac{kf}, which depends on the super-state $S_K^m$; see Eq. \eqref{eq1.3}. Such a filter predicts the continuous state associated with a super-state $S_k^m$, where $p(X_k|X_{k-1},S_{k-1}^m)$; and the posterior probability $p(X_k,S_k^m|Z_{k})$ is estimated using the current observation $Z_k$. 

Abnormalities can be seen as deviations between \ac{mjpf}'s predictions and the actual observed trajectories.
As a probabilistic filtering approach is considered, two main moments can be distinguished: $i)$ \textit{Prediction}: estimation of future states at a given time $k$. $ii)$ \textit{Update}: computation of the state's posterior probability based on the comparison between predicted states and new measurements. Accordingly, abnormality behaviors can be measured in the update phase, i.e., when predicted probabilities are far from observations. As it is well known, innovations in KFs are defined as:
\begin{equation}\label{eq4}
\epsilon_{k}^m = Z_k - H\hat{X}^m_{k|k-1},
\end{equation}
where $\epsilon_{k}^m$ is the innovation generated in the zone $m$ where the agent is located at a time $k$. $Z_k$ represents observed spatial data and $\hat{X}^m_{k|k-1}$ is the KF estimation of the agent's location at the future time instant $k$ calculated at the time $k-1$ by using Eq. \eqref{eq1.3}. Additionally, $H$ is the observation model that maps measurements into states, such that $Z_k = H X_k + v$ where $v \sim \mathcal{N}(0,R)$ and $R$ is the covariance observation noise.

Abnormalities are moments when a tracking system fails to predict subsequent observations. Consequently, we propose a weighted norm of innovations for detecting them, such that:
\begin{equation}\label{eq5}
\mathcal{Y}_k^m = \mathbf{d}(Z_k,H\hat{X}^m_{k|k-1}).
\end{equation}
In the \ac{mjpf}, the expression shown in Eq. \eqref{eq5} is computed for each particle and the median of such values is used as a global anomaly measurement of the filter. Further details about the implemented method can be found in \cite{fusion_valentina}.

\subsection{Private Layer: On-line testing GANs}
\label{subsubsec:Online}

Once the GANs hierarchy $\{{\cal H}_{\boldsymbol{C}^{m'}}\}$ is trained, it can be used for online prediction and anomaly detection purposes. This section describes the testing phase for state/label estimation and the proposed method for the detection of abnormalities.

\noindent{\emph{Label estimation}: }At testing time, we aim at estimating the state and detect possible abnormalities from the training set. More specifically, let ${D}^{0:{\cal I} \rightarrow {\cal O}}$ and ${D}^{0:{\cal O} \rightarrow {\cal I}}$ be the patch-based discriminators trained using the two channel-transformation tasks. Given a test frame ${\cal I}_k$ and its corresponding optical-flow image ${\cal O}_k$, we first produce the reconstructed ${\cal P}^{0:\cal O}_k$ and  ${\cal P}^{0:\cal I}_k$ using the first level generators ${G}^{0:{\cal I} \rightarrow {\cal O}}$ and ${G}^{0:{\cal O} \rightarrow {\cal I}}$, respectively. Then, the pairs of patch-based discriminators ${D}^{0:{\cal I} \rightarrow {\cal O}}$ and ${D}^{0:{\cal O} \rightarrow {\cal I}}$, are applied for the first and the second task, respectively. This operation results in two score maps for the observation: $D^{0:{\cal I} \rightarrow {\cal O}} ( {\cal O}_k,  {\cal I}_k)$ and $D^{0:{\cal O} \rightarrow {\cal I}} ( {\cal I}_k,  {\cal O}_k)$, and two score maps for the prediction (the reconstructed data): $D^{0:{\cal I} \rightarrow {\cal O}} ( {\cal P}^{0:\cal O}_k,  {\cal I}_k)$ and $D^{0:{\cal O} \rightarrow {\cal I}} ( {\cal P}^{0:\cal I}_k,  {\cal O}_k)$. In order to estimate the state, we used Eq. \eqref{eq:dist_gan} to generate the joint representation ${\cal X}_k = [{\cal X}^{\cal I}_k , {\cal X}^{\cal O}_k]$, where:

\begin{equation}\label{eq:state_gan}
\begin{multlined}
\quad{\cal X}^{\cal I}_k = D^{0:{\cal O} \rightarrow {\cal I}} ( {\cal I}_k,  {\cal O}_k) -  
D^{0:{\cal O} \rightarrow {\cal I}} ( {\cal P}^{\cal I},  {\cal O}_k) ,\\
{\cal X}^{\cal O}_k = D^{0:{\cal I} \rightarrow {\cal O}} ( {\cal O}_k,  {\cal I}_k) -  
D^{0:{\cal I} \rightarrow {\cal O}} ( {\cal P}^{\cal O}_k,  {\cal I}_k)\quad 
\end{multlined}
\end{equation}

For estimating the current super-state, we use ${\cal X}_k$ to find the closest SOM's detected prototypes. This procedure is repeated for all the models in the hierarchy $[{\cal H}_{\boldsymbol{C}^{m'}}]$.
This model can be seen as a switching model (see Fig. \ref{fig:dbn_pl}), where a hierarchy of GANs estimates the continuous space of the states whereas a hidden Markov model predicts the discrete space of the states. 

\noindent{\emph{Anomaly detection}: }Note that, a possible abnormality in the observation (e.g., an unusual object or an uncommon movement) corresponds to an outlier with respect to the data distribution learned by ${\cal N}^{{m':}{\cal I} \rightarrow {\cal O}}$ and ${\cal N}^{{m':}{\cal O} \rightarrow {\cal I}}$ during training. The presence of the anomaly, results in a low value of prediction score maps: $D^{{m':}{\cal O} \rightarrow {\cal I}} ( {\cal P}^{\cal I},  {\cal O}_k)$ and $D^{{m':}{\cal I} \rightarrow {\cal O}} ( {\cal P}^{\cal O}_k,  {\cal I}_k)$, but a high value of observation score maps: $D^{{m':}{\cal O} \rightarrow {\cal I}}_l ( {\cal I}_k,  {\cal O}_k)$ and $D^{{m':}{\cal I} \rightarrow {\cal O}} ( {\cal O}_k,  {\cal I}_k)$. 
Hence, to decide whether an observation is normal or abnormal, we calculate the average value of the \textit{innovations maps} introduced in Eq. \eqref{eq:state_gan} from both modalities. Therefore, the final abnormality measurement is defined as: 
\begin{equation}\label{eq:dist}
\theta_k = \overline{{\cal X}^{\cal I}_k} + \overline{{\cal X}^{\cal O}_k}
\end{equation}

The final representation of \ac{pl} for an observation ${\cal Z}_k = ({\cal I}_k , {\cal O}_k )$ consists of the computed $\theta_k$ and estimated super-state $C^{m'}_k$. We define an error threshold $\theta_{th} = \psi_{\boldsymbol{C}^{m'}}$ to detect the abnormal events: when all the levels in the hierarchy of GANs classify a sample as abnormal (e.g., dummy super-state) and the measurement $\theta_k$ is higher than $\theta_{th}$, the current measurement is considered as an abnormality. Note that the process is aligned to the one in the \ac{sl} layer, with the advantage that GANs deal with high multi-dimensional inputs and non-linear dynamic models at the continuous level. This complexity is required to analyze video data involved in \ac{pl}.


\section{Experimental Results}
\label{sec:res}
In this section, first, we introduce our collected dataset, then present the off-line training process for each layer of the self-awareness model, and finally, we demonstrate the results of online abnormality detection for the proposed test scenarios.  
\subsection{Experimental Dataset}
\label{sec:datasets}
In our experiments, an iCab vehicle \cite{Marin2016} driven by a human operator is used to collect the dataset.  We obtain the vehicle's positions mapped into Cartesian coordinates from the odometry manager~\cite{Marin2016}, as well as captured video footage from a first-person vision acquired by a built-in camera of the vehicle. 
For the SL, measurements of the iCab's positions and its flow components are considered. Furthermore, for the cross-modal GANs in the PL, we input the captured video frames and their corresponding optical-flow maps.

Accordingly, this work considers three situations (experiments): \emph{Scenario I)} or normal perimeter monitoring, where the vehicle follows a rectangular trajectory around a building (see Fig.~\ref{fig:subplans}-a). \emph{Scenario II)} or U-turn, where the vehicle performs a perimeter monitoring and faces a pedestrian, so it makes a U-turn to continue the task in the opposite direction (see Fig.~\ref{fig:subplans}-b). \emph{Scenario III)} or emergency stop, where the vehicle encounters pedestrians crossing its path and stops until the pedestrian leaves its field of view (see Fig.~\ref{fig:subplans}-c).


Situations II and III contain deviations from the perimeter monitoring task. 
When an observation falls outside the super-state, as the learned models are not valid, a {\em dummy neuron} is used to represent the new experience, and random filter where $U = 0_{2,1}$ in Eq.~\eqref{eq1.3} is employed for prediction purposes. 
\subsection{Training SL and PL}
The two layers of the proposed self-awareness model, including the \ac{sl} (modeled by a \ac{mjpf}) and the \ac{pl} (modeled as a hierarchy of GANs), are able to learn the normality. In our experiments, normal behaviors are performed as Scenario I (Fig.~\ref{fig:subplans}-a). Such a scenario is used as a training set to learn both models (\ac{sl} and \ac{pl}). For each layer, we use different observations. Accordingly, \ac{sl} uses positional information while \ac{pl} utilizes first-person visual data. In the rest of this section, we show the training process and output of each layer.

\begin{figure*}[t]
\begin{minipage}{.5\textwidth}
  \centering
  \includegraphics[width=.95\linewidth]{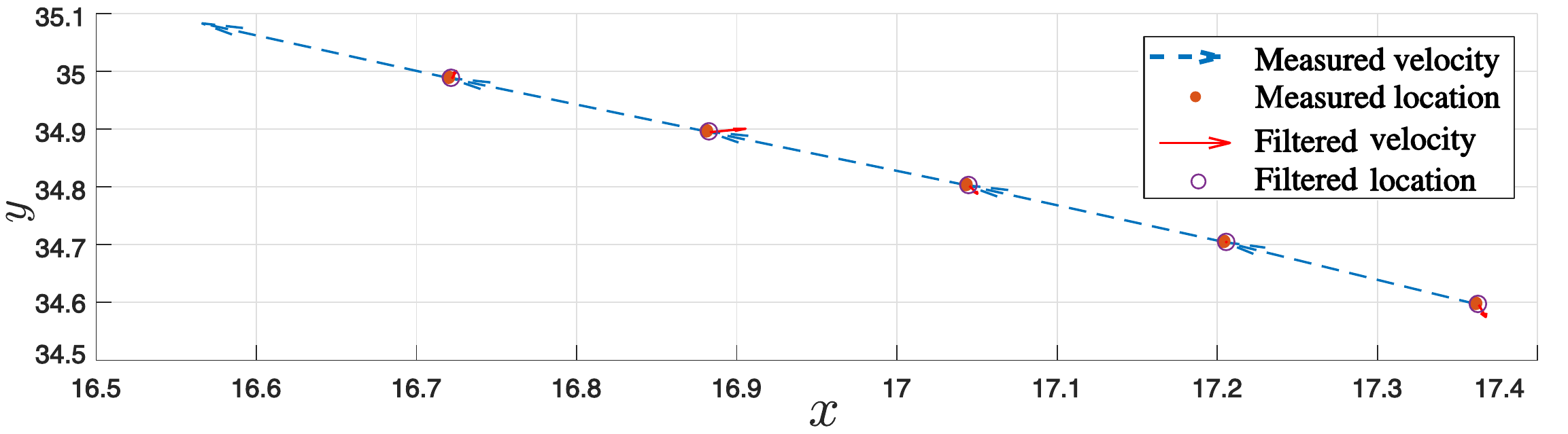}
  \caption{U-KF's estimations}
  \label{fig:sfig1}
\end{minipage}%
\begin{minipage}{.5\textwidth}
  \centering
  \includegraphics[width=.95\linewidth]{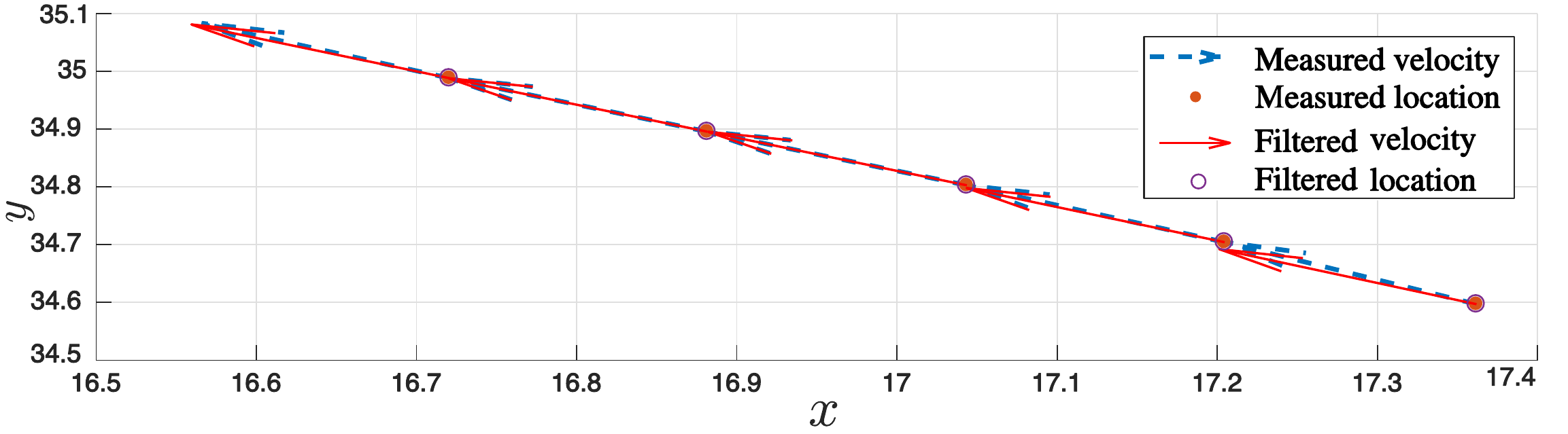}
  \caption{Learned filter's estimations}
  \label{fig:sfig2}
\end{minipage}

\begin{minipage}{.5\textwidth}
  \centering
  \includegraphics[width=.95\linewidth]{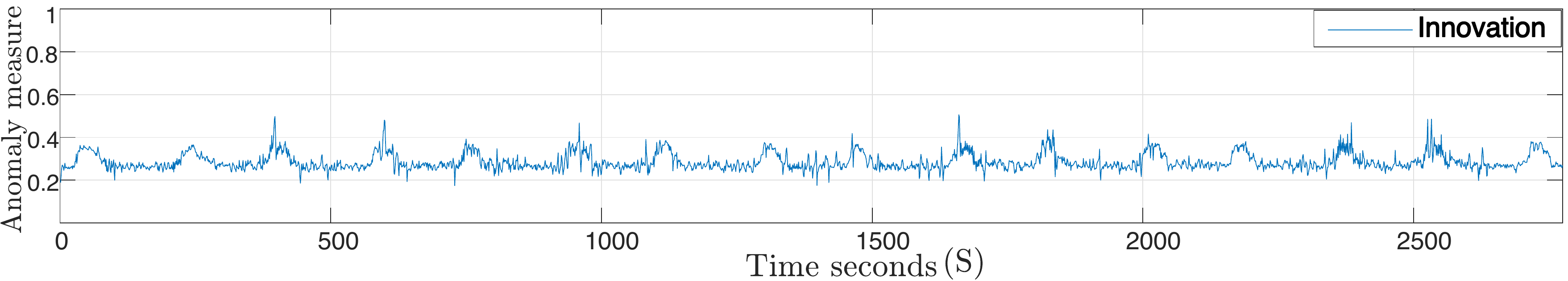}
   \caption{U-KF's innovations}
  \label{fig:sfig3}
\end{minipage}%
\begin{minipage}{.5\textwidth}
  \centering
  \includegraphics[width=.95\linewidth]{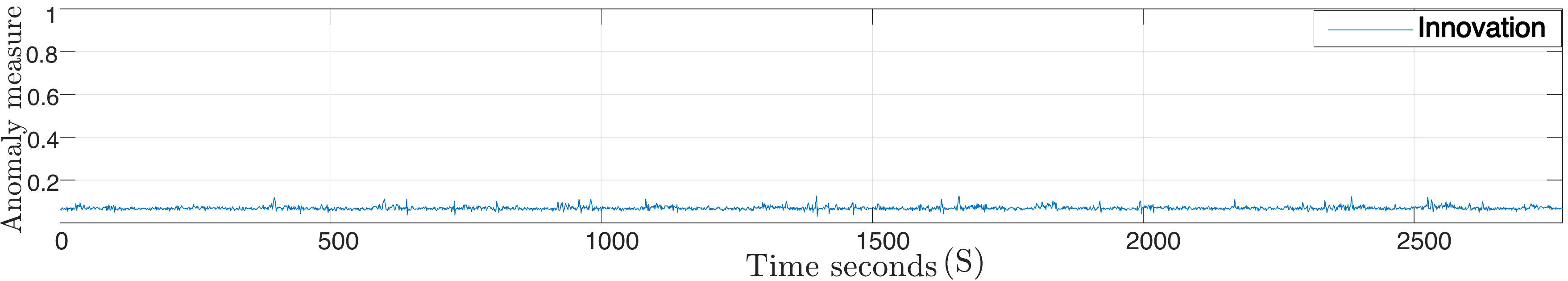}
  \caption{Learned filter's innovations}
  \label{fig:sfig4}
\end{minipage}
\caption{SL state estimation: (a) and (b) correspond to the estimations made by the U-KF and the learned filter respectively. Abnormality signals in SL: (c) and (d) show the innovations generated by the U-KF and the learned filter respectively.}
\label{fig:ukf_training}
\end{figure*}
\begin{figure*}[t]
	\centering
	(a)\includegraphics[width=0.30\linewidth,height=0.07\linewidth]{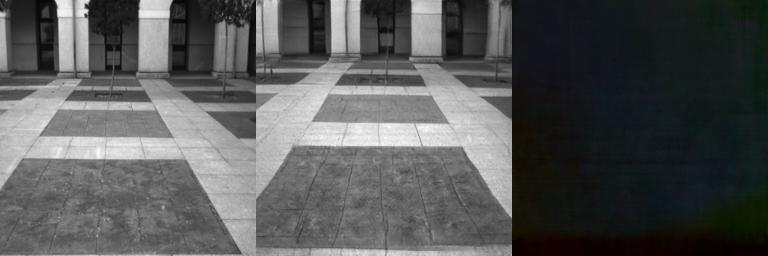}
	(c)\includegraphics[width=0.30\linewidth,height=0.07\linewidth]{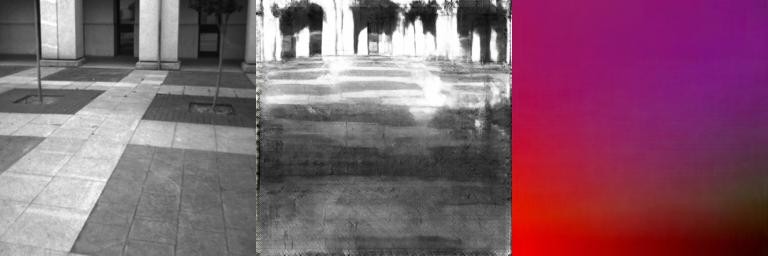}
	(e)\includegraphics[width=0.30\linewidth,height=0.07\linewidth]{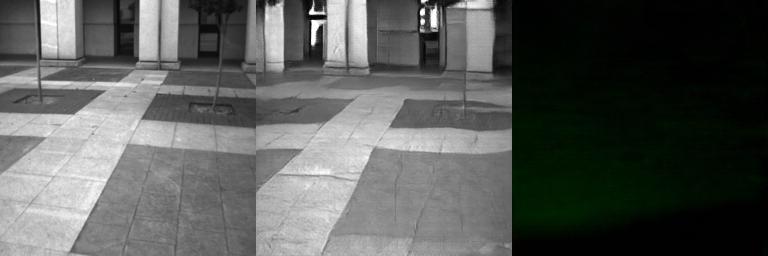}
	(b)\includegraphics[width=0.30\linewidth,height=0.07\linewidth]{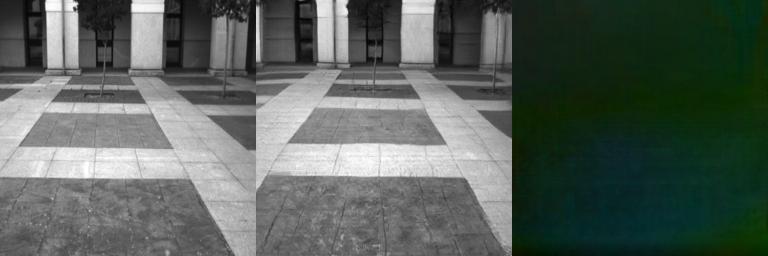}
	(d)\includegraphics[width=0.30\linewidth,height=0.07\linewidth]{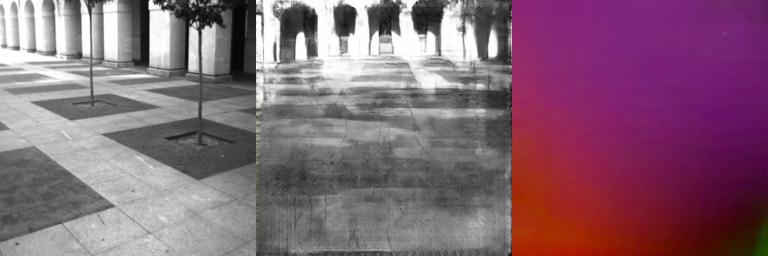}
	(f)\includegraphics[width=0.30\linewidth,height=0.07\linewidth]{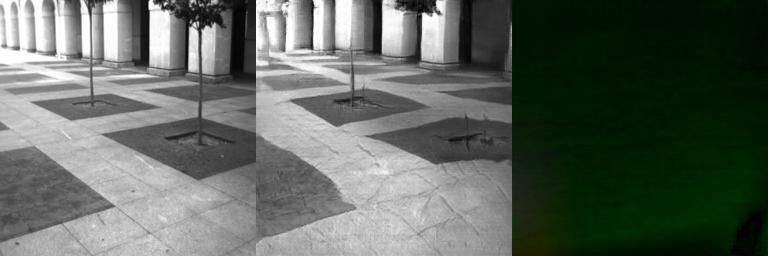}
	\caption{ PL state estimation: (a) and (b) are the estimated video frame and optical-flow motion map from the \emph{reference GAN} while the vehicle moves toward a linear path. (c) and (d) are the estimations from the \emph{reference GAN} while the vehicle curving, (e) and (f) are the same samples predicted by the hierarchy of GANs after training. Each triplet image contains the ground truth observation (left), the predicted frame (center), and the difference between prediction optical-flow map and the ground truth where black pixels indicate a high accuracy at the prediction stage.}
	\label{fig:gan_training_estimation}
\end{figure*}
\noindent{\textbf{Initial model in SL.}}
As we discussed in Sec. \ref{sec:method}, the reference filter for the shared layer is a U-KF; see Fig. \ref{fig:learning_bd}-(a.1), which assumes the simple condition of equilibrium in which the agent is not moving. A sequence of state estimation samples of U-KF is shown in Fig. \ref{fig:sfig1}. In this figure, U-KF always predicts the agent will stay in a still position, including a negligible perturbation error; see small velocity vectors in red.  However, this assumption is not true in the observed data; see large velocity vectors in blue in Fig. \ref{fig:sfig1} and leads to large innovation values such as shown in Fig. \ref{fig:sfig3}. 

\noindent{\textbf{Training incremental models in SL.}}
By applying the reference filter U-KF over the training data, a set of innovation values can be obtained. This set of innovations is plotted in Fig. \ref{fig:sfig3}, and it is used in the next training iteration to learn new filters incrementally as described in Sec. \ref{subsec:learn_SL} (see Fig. \ref{fig:learning_bd}-(d.1)). The new set of MKFs encode models that describe new conditions of equilibrium and whose estimations are more accurate estimations than the ones produced by the  U-KF when dealing again with similar abnormal situations, compare blue and red arrows in \ref{fig:sfig2}. It can be seen that predictions are close to the observations, producing low errors, i.e., low innovation values, as shown in Fig. \ref{fig:sfig4}.

\begin{figure}[htb]
\begin{minipage}[b]{\linewidth}
  \centering
  \setlength\tabcolsep{0pt}
  \begin{tabular}{p{0.05\linewidth} p{0.95\linewidth}}
     (a)& \hspace{0.1cm} \includegraphics[width=.97\linewidth,height=0.4cm]{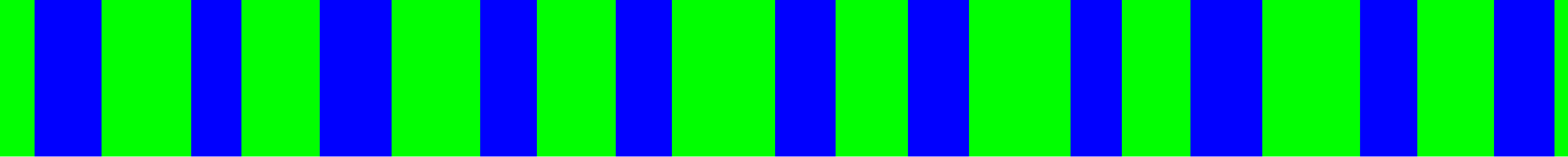}\\
     (b)&  \includegraphics[width=\linewidth,trim={3.9cm 0 3.7cm 0},clip]{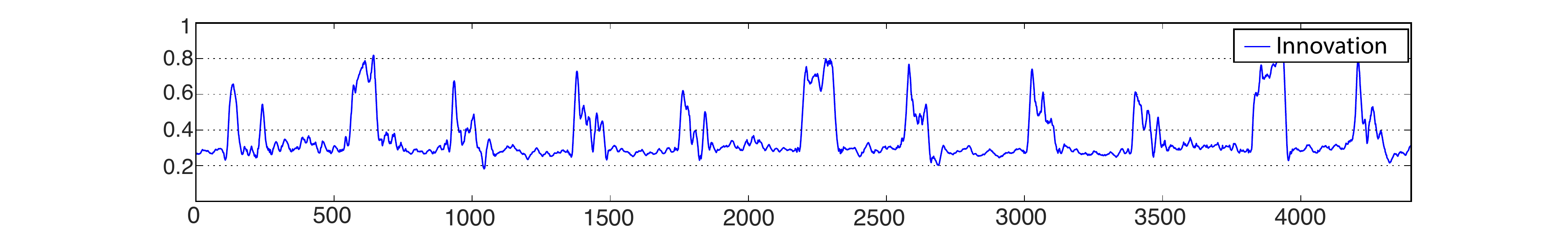}\\
     (c)&  \includegraphics[width=\linewidth,trim={3.9cm 0 3.7cm 0},clip]{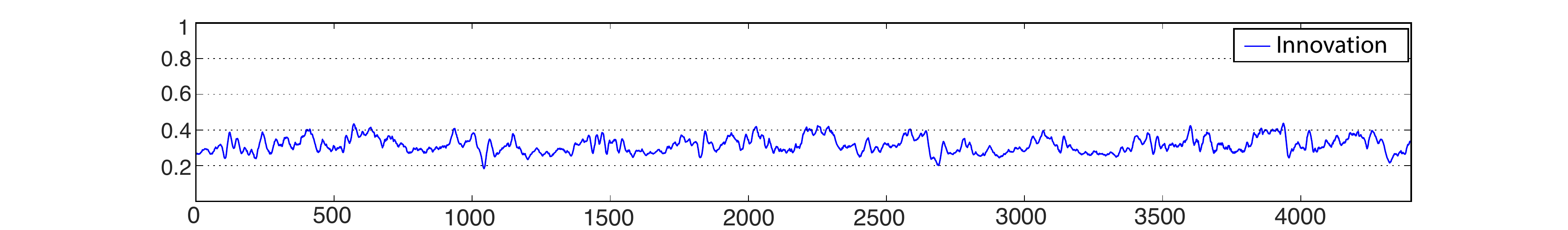}\\
  \end{tabular}
\end{minipage}
\caption{Training hierarchy of GANs: (a)  ground truth labels, the green background means the vehicle moves on a linear path, while the blue bars indicate curving. (b) and (c) show the signal of the averaged score distance values between prediction and observation (innovation) for the first level of GANs and the hierarchy of GANs, respectively. Horizontal and vertical axes encode time and innovation values correspondingly.}
\label{fig:train}
\end{figure}

\noindent{\textbf{Initial model in PL.}}
As mentioned in Sec. \ref{subsec:learn_PL}, constructing the GAN hierarchical model is done based on the distance of discriminators scores between the predictions and the real observations. The first level of GANs (\emph{reference GANs}) is trained on a selected subset of normal samples from \emph{perimeter monitoring} sequence. This subset represents the captured sequences while the vehicle moves on a linear path in a normal situation, i.e., when the road is empty and the vehicle moves linearly. The hypothesis is that this subset only represents one of the motion distributions and appearance in a highly diverse data condition. As a result, when the pair of \emph{reference GANs} detects an abnormality in the corresponding set on which is trained, the corresponding observations can be considered as outliers. This hypothesis is confirmed by testing the \emph{reference GANs} over the entire sequences of \emph{perimeter monitoring}, and observing the discriminators' scores distances between the prediction and the observation. 
Fig. \ref{fig:train} shows the results of training \emph{reference GANs}. Our hypothesis concerning the complexity of distributions is confirmed in Fig. \ref{fig:train} (a), where the test is performed using only the \emph{reference GANs}. 

\noindent{\textbf{Training incremental models in PL.}}
As shown in Fig. \ref{fig:train}, \emph{reference GANs} can predict/detect the linear path (white background area) perfectly. On the other hand, when the vehicle curves (green bars), the system fails and recognizes curving as an abnormal event. This means, the \emph{reference GAN} discriminators' scores distances between the prediction and the observation (abnormality signal) are higher over the curving areas, which was expected. However, after collecting this set of abnormal data and training the second level of GANs, the newly learned models facilitate recognizing the entire training sequence as normal. 

The estimation of optical-flow and frame for each level (iteration) of the training process is shown in Fig. \ref{fig:gan_training_estimation}. For each case, an image triplet is shown, where the left image is the ground truth observed frame, the central image shows the predicted frame, and the right image is the difference between the observed optical-flow motion map and the predicted optical-flow. The lower the distance between predicted motion and observation, the blacker (values are near to ``$0$") the right image. In this figure, (a), (b), (c), and (d) show the output estimation for the initial GANs. As can be seen, straight motions displayed in (a) and (b) are correctly estimated; see the low error, i.e., black pixels, in the right frames of their triplets.
Nonetheless, the initial model is unable to predict curve motions shown in (c) and (d), see the high error, i.e., colorful pixels, in the right frames of their triplets. Fig. \ref{fig:gan_training_estimation} (e) and (f) show the estimation from the hierarchy of GANs after a full training phase in case of curving. It can be observed this time how the GANs estimate the curving motion with high accuracy, see the number of black pixels in the triplet's right frames. 

As the \emph{reference GANs} are trained on the reference situation which is the linear movements (see Fig. \ref{fig:learning_bd}-(a.2)), therefore is it expected to having a good estimation while the vehicle moving linearly (Fig. \ref{fig:gan_training_estimation}-a-b). However, this filter fails to estimate curves (Fig. \ref{fig:gan_training_estimation}-c and d). The incremental nature of the proposed method is demonstrated in Fig. \ref{fig:learning_bd}-(d.2), where low estimation errors are obtained by using the second level GANs for predicting curve motions, see \ref{fig:gan_training_estimation}-e-f. 
 
\begin{table*}[]
	\centering
	\begin{tabular}{p{1.55cm}p{1.5cm}p{0.8cm}p{2.3cm}p{1cm}p{1cm}p{1.5cm}p{2cm}p{1.5cm}}
		\hline
		Zone \#         & zone 1                                                                                  & zone 2         & zone 3                                                         & zone 4   & zone 5 & zone 6  & zone 7 & zone 8  \\ \hline
		SL ($\{S_k^m\}$) & 3-11, 17-23, 31-35, 45-47 & 12, 13, 24, 26 & 25, 38, 39, 49-52, 62-65, 76-78, 90, 91 & 103, 104 &  111-115    & 93, 105-107        &  28, 40-43, 53-56, 66-68, 79-81, 92      & 1, 2, 14-16, 27        \\
		PL ($\{C_k^{m'}\}$) & 1-3  & 4-7        & 1-3  & 4-7  & 1-3  & 4-7 & 1-3  & 4-7 \\ \hline
	\end{tabular}
	
	\caption{List of corresponding detected super-states from PL and SL for the normal scenario: for each zone, the number of color-coded super-states sequences from PL ($\{C_k^{m'}\}$) and SL ($\{S_k^m\}$) are shown.}
	\label{tab:pri_ss}
\end{table*}

\begin{figure}[t]
	\centering
	\includegraphics[width=0.8\linewidth]{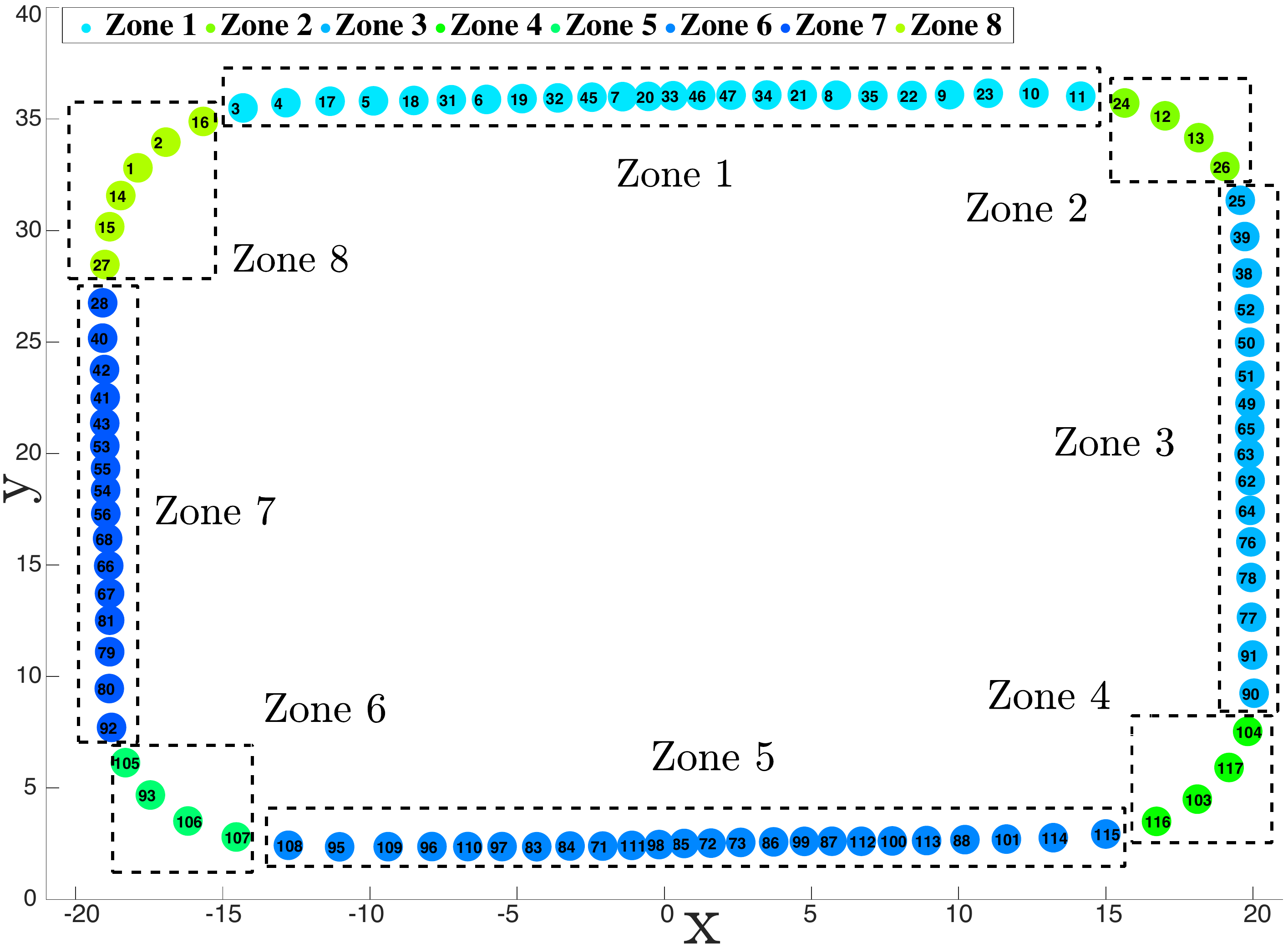}
	\caption{Color-coded zones from SL and PL.}
		\label{fig:zones}
\end{figure}

\subsection{Representation of normality in SL and PL}
\begin{figure}[t]
	\begingroup
\setlength{\tabcolsep}{1pt} 
\renewcommand{\arraystretch}{0.5} 
\begin{tabular}{m{0.45cm}p{\linewidth}}
    \scriptsize{(a)} & \includegraphics[width=0.91\linewidth,height=0.4cm]{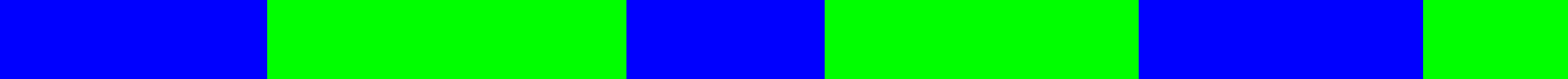} \\
    \scriptsize{(b)} & \includegraphics[width=0.91\linewidth,height=0.4cm]{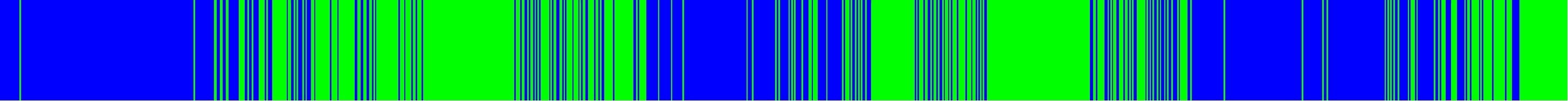}\\
    \scriptsize{(c)} & \includegraphics[width=0.91\linewidth,height=0.4cm]{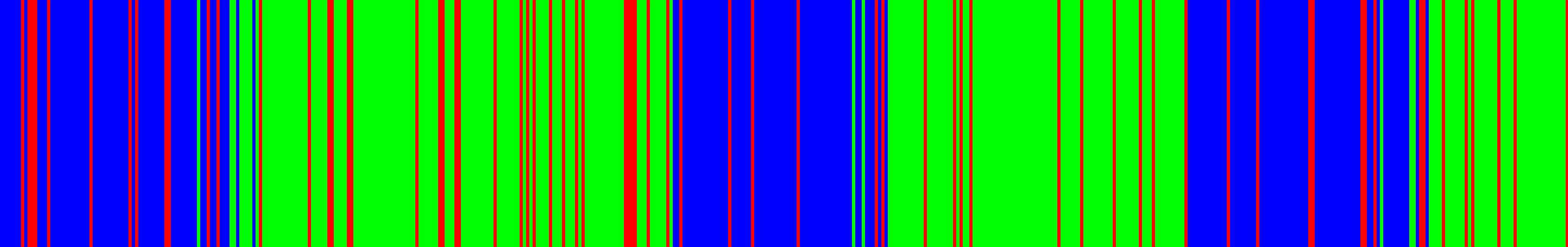}
\end{tabular}
\endgroup
	\centerline{\scriptsize{(d)}\includegraphics[width=0.96\linewidth,trim={3.9cm 0 3.2cm 0},clip]{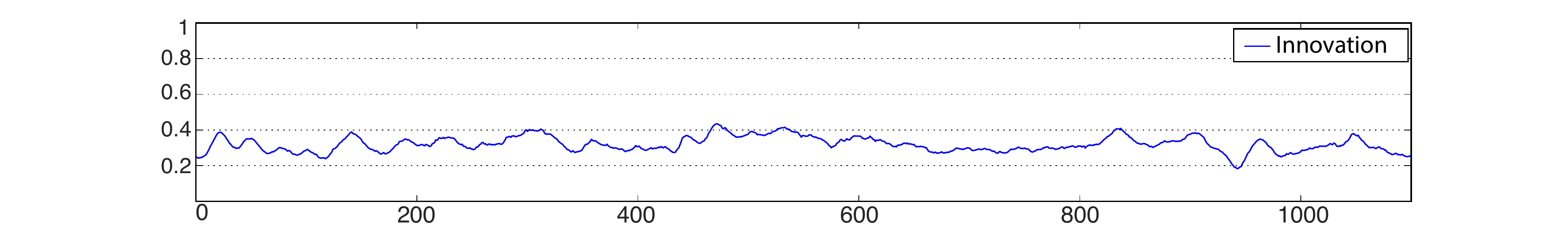}}
	\centerline{\scriptsize{(e)}\includegraphics[width=0.96\linewidth,trim={4.8cm 0 3.5cm 0},clip]{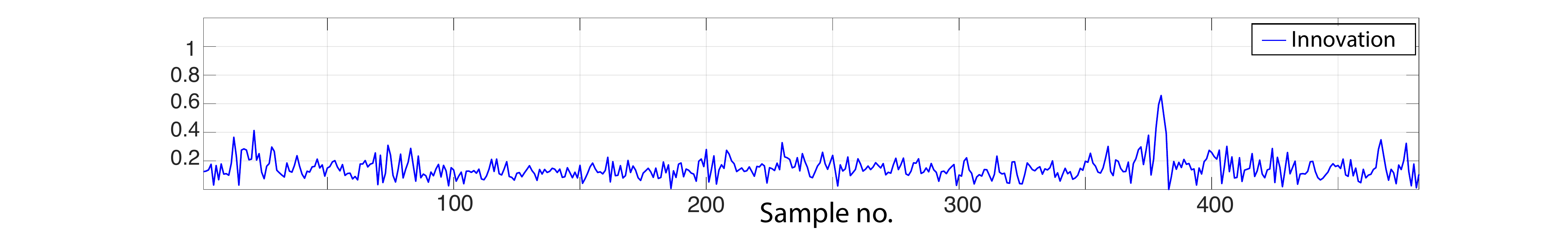}}
	\caption{Normality representations of \ac{pl} and \ac{sl}: (a) shows the ground truth labels, green and blue bars represent linear and curve motions respectively. (b) and (c) show color-coded super-states sequences $\{C_k^{m'}\}$ and $\{S_k^m\}$ respectively. They are highly correlated with the agent's real status (a). Images (d) and (e) show the abnormality signals from \ac{pl} and \ac{sl}, respectively. The horizontal axes in (d) and (e) are the sample numbers, and the vertical axes show the abnormality signals.}
	\label{fig:pri_states}
\end{figure}

After training \ac{sl} and \ac{pl} over the training sequence, to evaluate the final learned models, we select a period of the normal perimeter monitoring task (see Fig. \ref{fig:subplans}-a) as a test scenario. As reviewed in Sec. \ref{sec:method} both \ac{sl} and \ac{pl}, represent their situation awareness by a set of super-states following and abnormality signals. This set of results for \ac{pl} and \ac{sl} is shown in Fig. \ref{fig:pri_states}, which simply visualizes the learned normality representations. The ground truth label is shown in Fig. \ref{fig:pri_states}-a, and the color-coded detected super states from \ac{pl} $\{{C^{m'}}\}$ and \ac{sl} $\{S^m\}$ are illustrated in Fig. \ref{fig:pri_states}-b and Fig. \ref{fig:pri_states}-c, respectively. It clearly shows that the pattern of super-states is repetitive and highly-correlated with the ground truth. It also shows a strong correlation between the sequence of \ac{pl} and \ac{sl} super-states.
 
The study of the cross-correlation between the \ac{sl} and \ac{pl} is beyond the scope of this paper. Nonetheless, it is also interesting to demonstrate such a relation. For showing the correlation of both learned models (\ac{sl} and \ac{pl}), we divided the environment into eight meaningful zones, including curves and linear paths. This semantic partitioning of state-space is shown in Fig.~\ref{fig:zones}. For the training scenario (normal situation), the color-coded super-states of \ac{sl} and \ac{pl} are visualized over the environment plane. As shown in Table. \ref{tab:pri_ss}, for the \ac{sl} we detect 115 super-states, each of them corresponding to an individual filter. For \ac{pl}, we have detected seven different super-states, corresponding to trained cross-modal GANs.

Note that in Fig. \ref{fig:pri_states} (d) and (e), abnormality signals are stable for both \ac{pl} and \ac{sl} due to the presence of normal data. In the case of abnormalities, peaks (high local values) are expected over the signals. To study such abnormal situations, we apply the trained models over unseen test sequences. Two different scenarios are selected where the agent performs new maneuver actions in order to solve the abnormal situation.

\subsection{Abnormality detection in dynamic data series}
In this phase, we perform an online testing setup to evaluate the performance of our models. This procedure for the \ac{sl} is shown in Fig. \ref{fig:learning_bd}-(b.1), (c.1) and (d.1). For the PL, it is shown in Fig. \ref{fig:learning_bd}-(b.1), (c.1) and (d.1).
\begin{figure*}[t]
	\begin{center}
		\begin{minipage}[t]{0.32\textwidth}
			\centering
			\includegraphics[width=\textwidth,trim={0.89cm 0.18cm 1.25cm 0.68cm},clip]{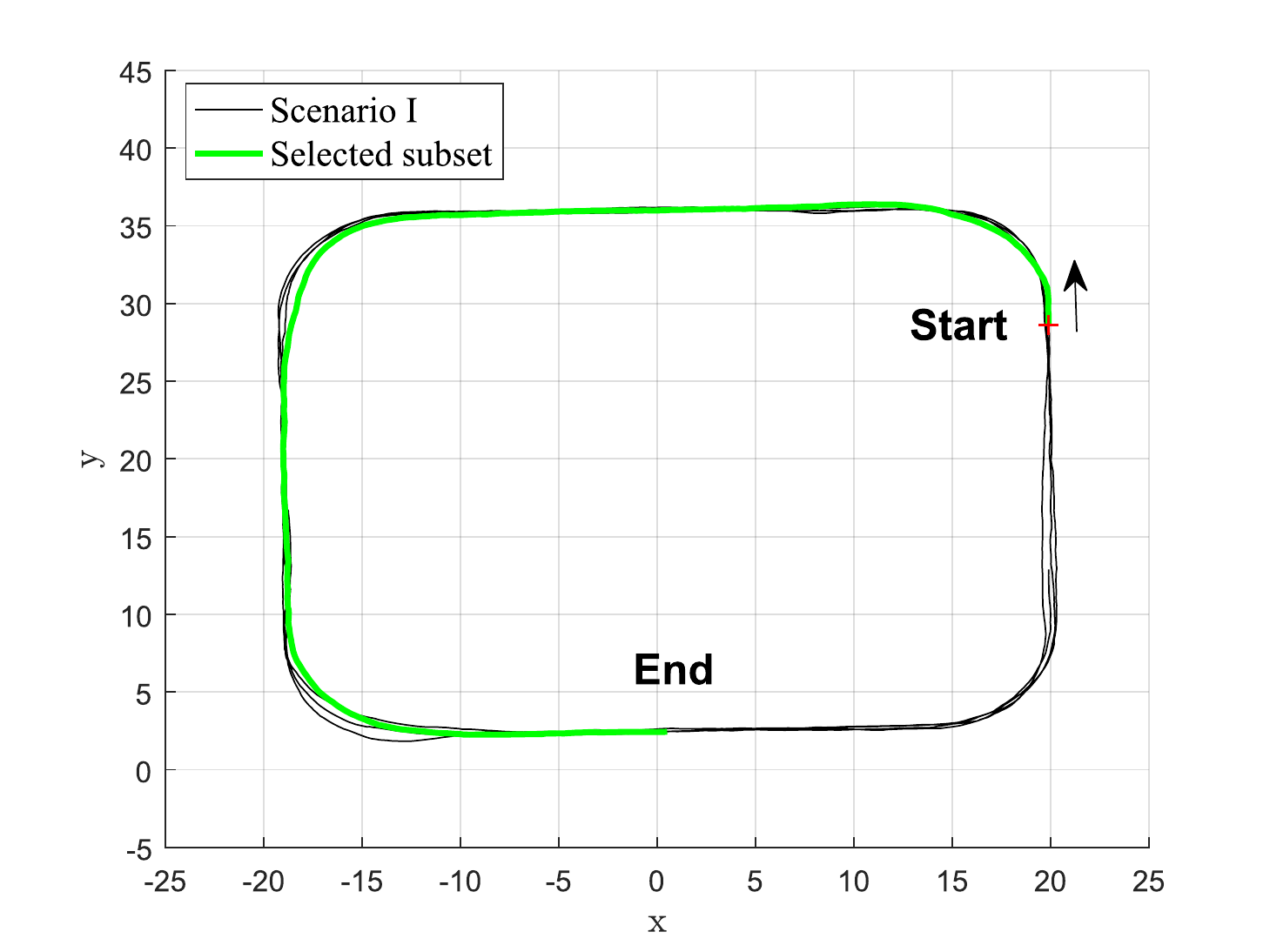}
			\scriptsize{(a) perimeter monitoring}
		\end{minipage}
		\begin{minipage}[t]{0.32\textwidth}
			\centering
			\includegraphics[width=\textwidth,trim={0.9cm 0.2cm 1.25cm 0.7cm},clip]{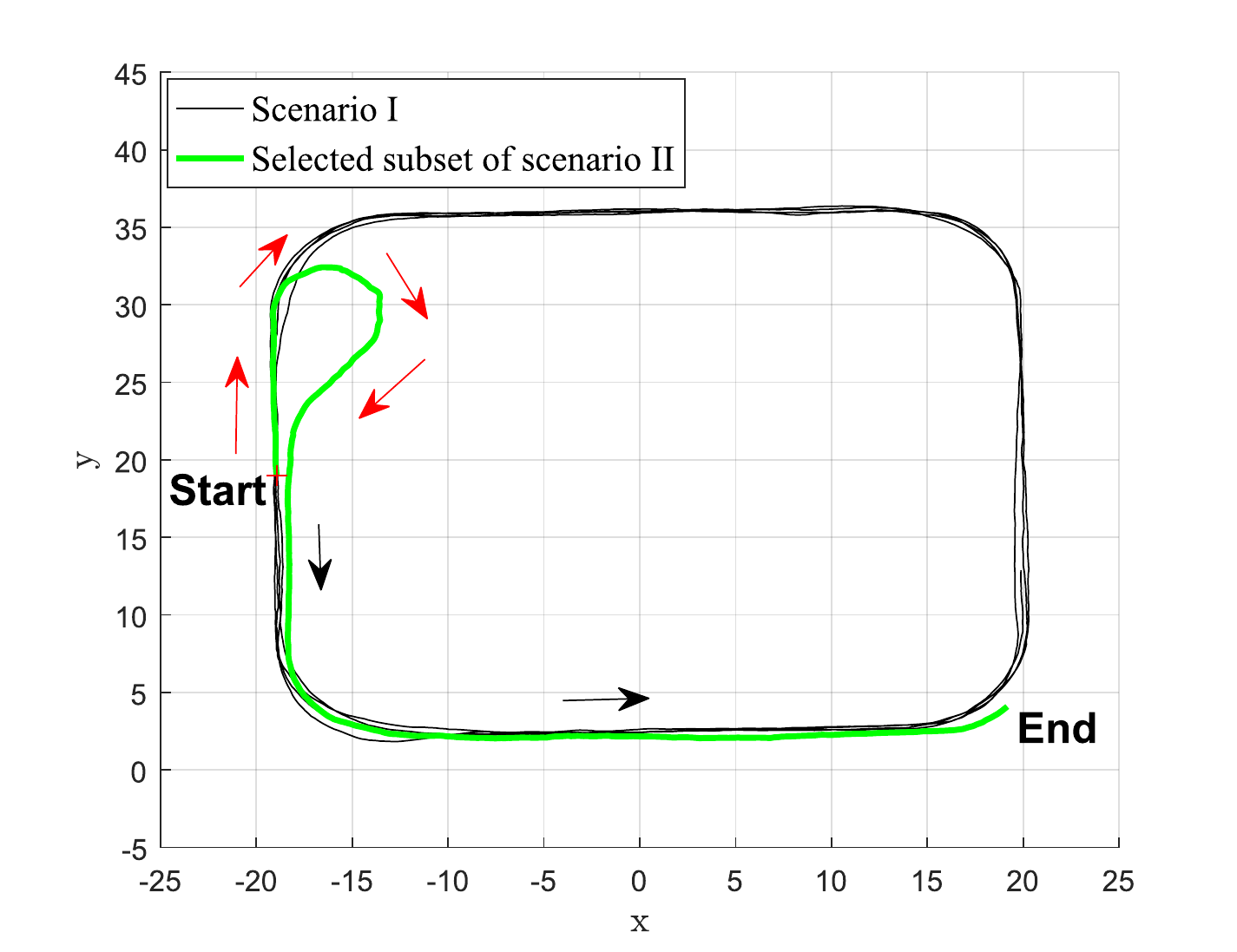}
			\scriptsize{(b) U-turn}
		\end{minipage}
		\begin{minipage}[t]{0.32\textwidth}
			\centering
			\includegraphics[width=\textwidth,trim={0.9cm 0.2cm 1.25cm 0.7cm},clip]{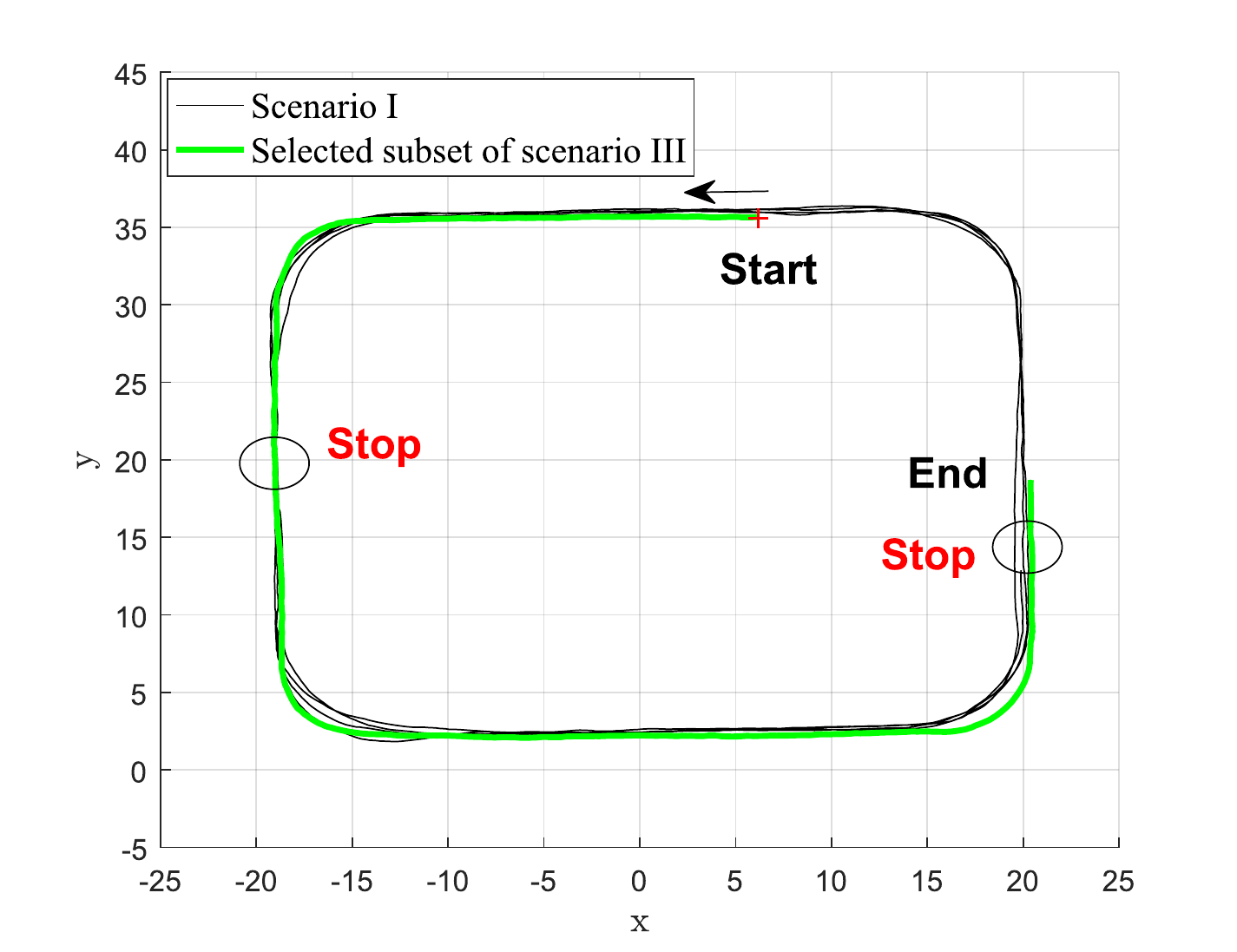}
			\scriptsize{(c) emergency stop}
			\label{fig:Stop}
		\end{minipage}
	\end{center}
	\caption{Sub-sequence examples from testing scenarios reported in the experimental results.}
	\label{fig:subplans}
\end{figure*} 

\begin{figure}[htb]
\begingroup
\setlength{\tabcolsep}{1pt} 
\renewcommand{\arraystretch}{0.5} 
\begin{tabular}{m{0.45cm}p{\linewidth}}
    \scriptsize{(a)} & \includegraphics[width=0.91\linewidth,height=0.4cm]{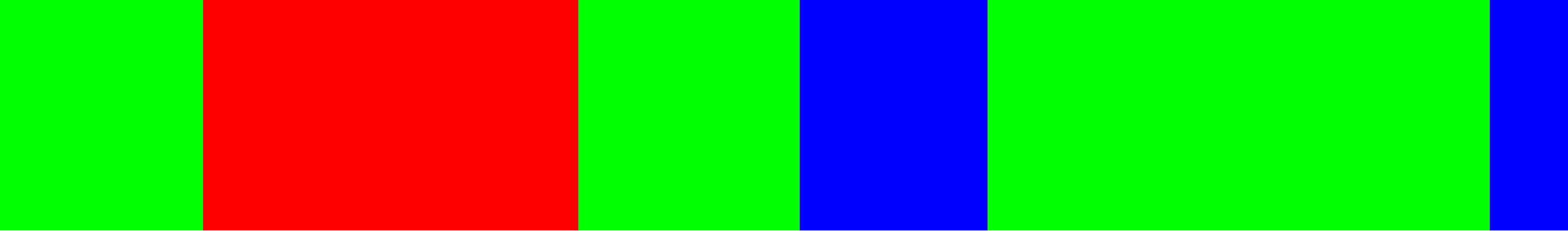} \\
    \scriptsize{(b)} & \includegraphics[width=0.91\linewidth,height=0.4cm]{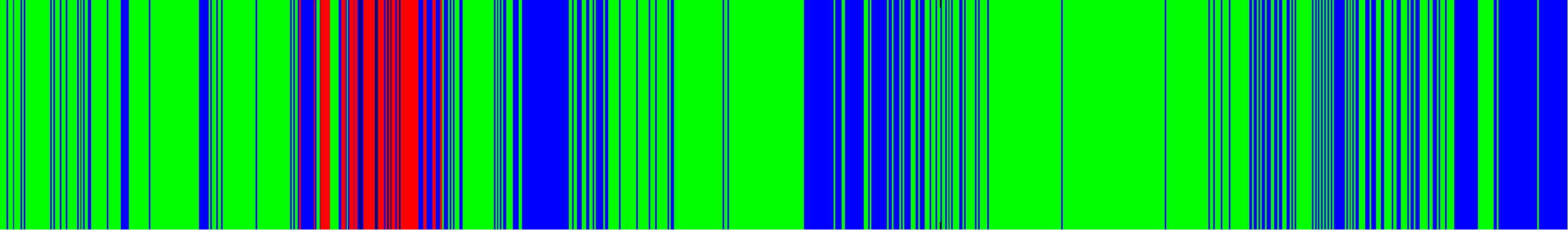}\\
    \scriptsize{(c)} & \includegraphics[width=0.91\linewidth,height=0.4cm]{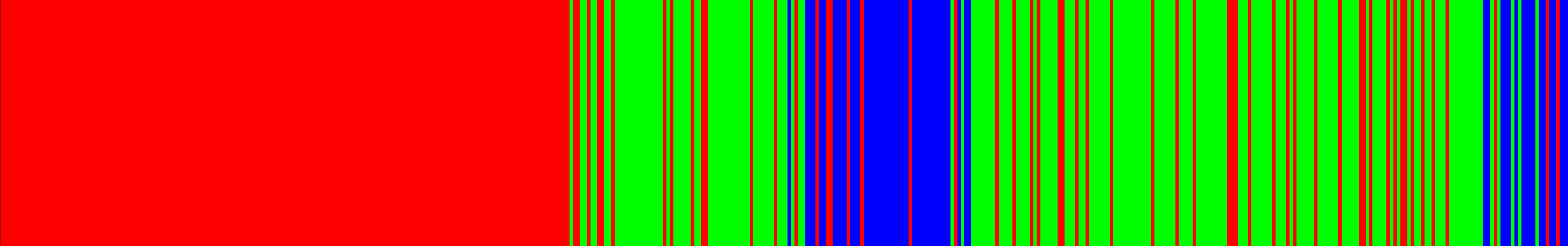}
\end{tabular}
\endgroup

	\centerline{\scriptsize{(d)}\includegraphics[width=0.96\linewidth,trim={3.9cm 0 3.2cm 0},clip]{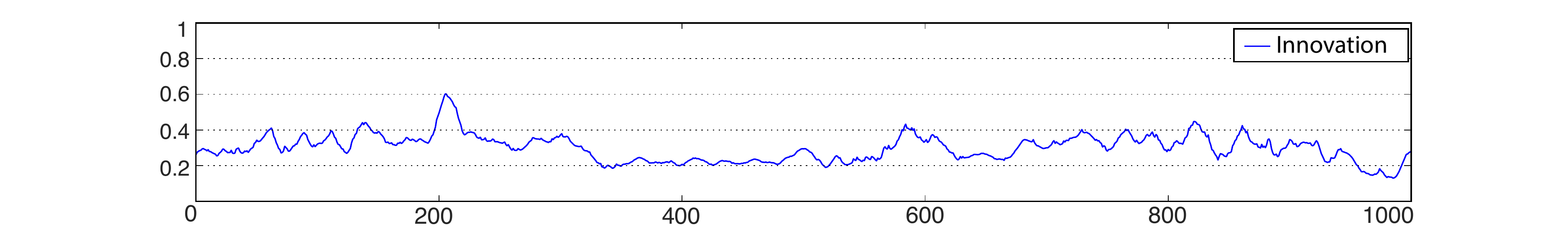}}
	\centerline{\scriptsize{(e)}\includegraphics[width=0.96\linewidth,trim={4.7cm 0 3.5cm 0},clip]{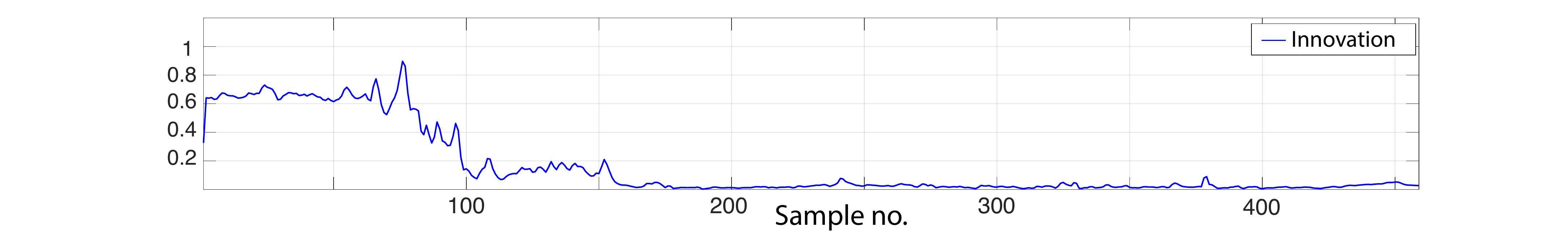}}
	\caption{Abnormality in the U-turn avoiding scenario: (a) ground truth labels. (b) and (c) color-coded transition of states $\{C_k^{m'}\}$ and $\{S_k^m\}$, respectively. (d) and (e) generated abnormality signal (innovation) from  PL and SL, respectively. The horizontal axis represents the sample number, and the vertical axis shows the innovation values (abnormality signal).}
	\label{fig:testushape}
\end{figure}

\noindent{\textbf{Avoiding a pedestrian by a U-turn action}:} In this scenario, which is illustrated in Fig.~\ref{fig:subplans}-b, the vehicle performs a U-turn avoidance maneuver when encounters a static pedestrian and then continues the standard monitoring in the opposite sense. The goal is to detect the abnormality, consisting of the pedestrian presence, which leads to unexpected agent's actions when compared to the learned normality during the perimeter monitoring. Fig. \ref{fig:testushape} shows the result of anomaly detection from \ac{pl} and \ac{sl}. The results are related to the highlighted time slice of the testing scenario II (Fig. \ref{fig:subplans}-b).

In Fig. \ref{fig:testushape}-a, the green background represents a vehicle's linear path, the blue bars indicate curving, and red bars show the presence of an abnormal situation (which corresponds to the static pedestrian). The abnormality area starts at first sight of the pedestrian, and it continues until the avoiding maneuver finishes (end of U-turn). Similarly, the sequences of states $\{C_k^{m'}\}$ and $\{S_k^m\}$ in Fig. \ref{fig:testushape}-(b,c), follow the same pattern. While the situation is normal, the super-states repeat the expected normal pattern, but as soon as the abnormality begins, the super-state patterns change in both \ac{pl} and \ac{sl} (e.g., dummy super-states). Furthermore, in the abnormal super-states, the abnormality signals present higher values.  
The abnormality signal (innovations)generated by \ac{sl} is shown in Fig. \ref{fig:testushape}-e. The abnormality produced by the vehicle is higher while it moves through the path, which is indicated by red arrows in Fig. \ref{fig:subplans}-b. This is caused by the observations that are outside the domain where super-states are trained. Namely, during the training phase, such a state-space configuration is never observed. Additionally, \ac{kf}'s innovations become higher in the same time interval due to the opposite velocity compared with the normal behavior of the model.

\begin{figure}[t]
	\centerline{\small{(a)}\includegraphics[width=0.95\linewidth,height=2.2cm]{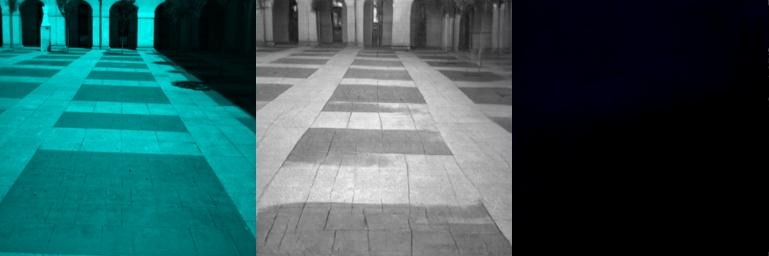}}
	\centerline{\small{(b)}\includegraphics[width=0.95\linewidth,height=2.2cm]{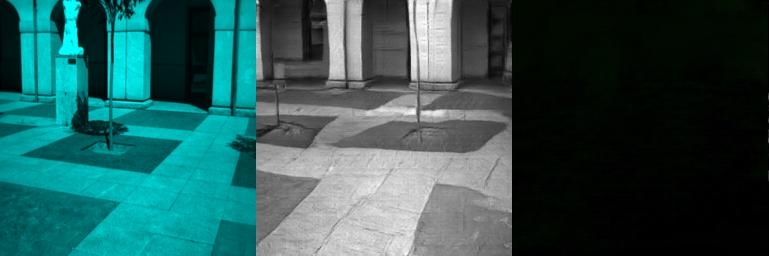}}
	\centerline{\small{(c)}\includegraphics[width=0.95\linewidth,height=2.2cm]{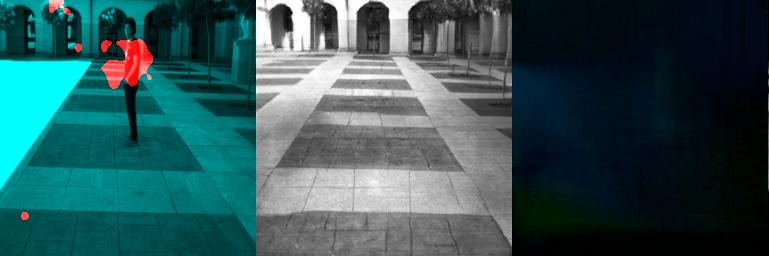}}
	\centerline{\small{(d)}\includegraphics[width=0.95\linewidth,height=2.2cm]{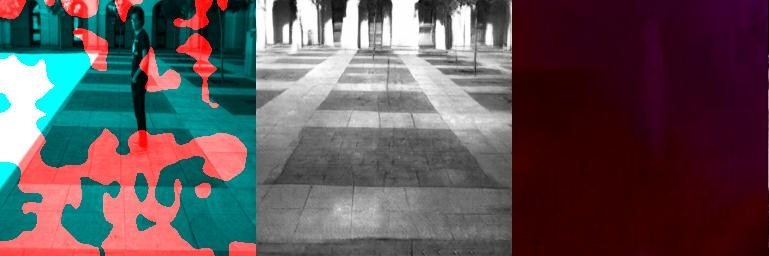}}
	\centerline{\small{(e)}\includegraphics[width=0.95\linewidth,height=2.2cm]{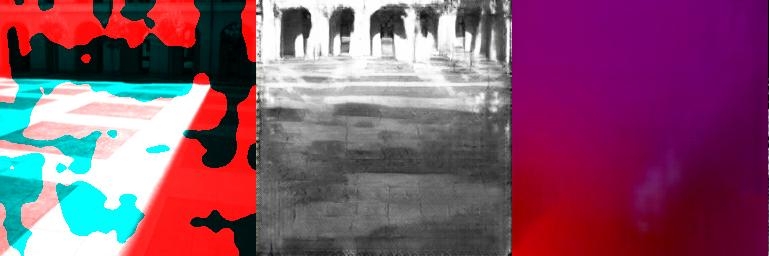}}
	\caption{Visualization of abnormality: the first column shows the localization over the original frame, the second column is the predicted frame, and the last column shows the pixel-by-pixel distance over the optical-flow maps. (a) moving linear, (b) curving, (c) first observation of the pedestrian, (d) and (e) performing the avoiding action.}
	\label{fig:visavoiding}
\end{figure}

The abnormality signal generated by \ac{pl}, Fig. \ref{fig:testushape}-d, is computed by averaging over the distance maps between prediction and observation score maps: when an abnormality arises, the proposed measure does not undergo large changes since a local abnormality (see Fig. \ref{fig:visavoiding}-c) can not change the average value significantly. However, as soon as it is observed a full sight of the pedestrian, and the vehicle starts performing the avoidance maneuver, the abnormality signal becomes higher since both observed appearance and action represent unknown situations. This situation is shown in Fig. \ref{fig:visavoiding}-(d,e). As soon as the agent back to the known situation (e.g., curving), the abnormality signal becomes lower.

\begin{figure}[t]
\begingroup
\setlength{\tabcolsep}{1pt} 
\renewcommand{\arraystretch}{0.5} 
	\begin{tabular}{m{0.45cm}p{\linewidth}}
    \scriptsize{(a)} & \includegraphics[width=0.91 \linewidth,height=0.4cm]{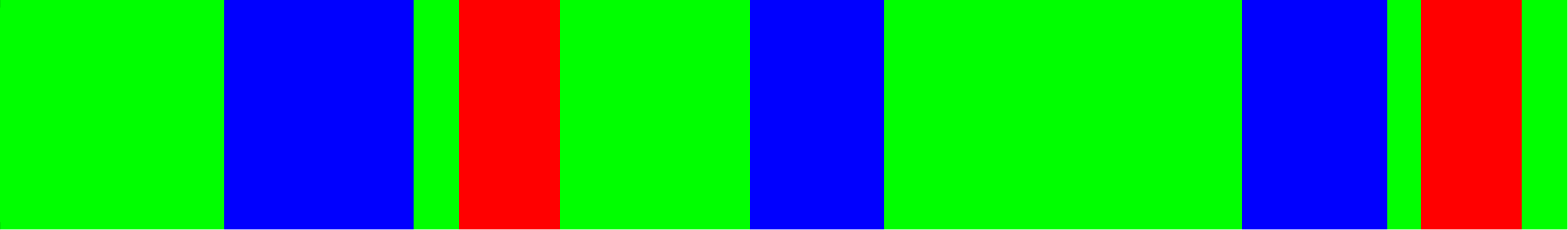} \\
    \scriptsize{(b)} & \includegraphics[width=0.91\linewidth,height=0.4cm]{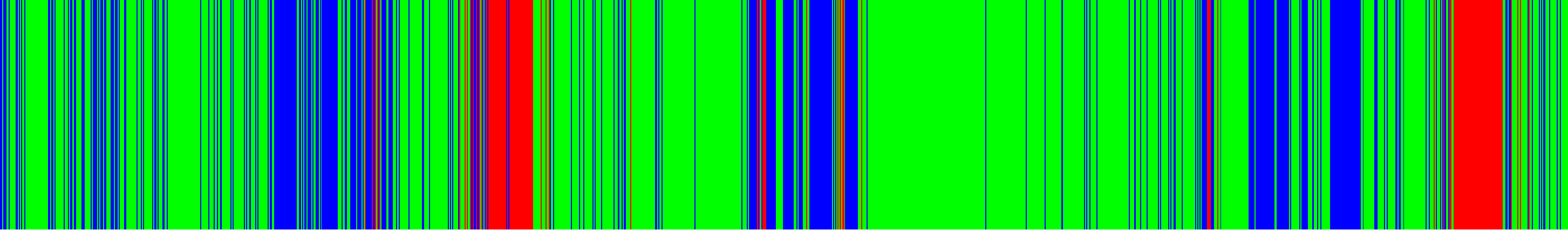}\\
    \scriptsize{(c)} & \includegraphics[width=0.91\linewidth,height=0.4cm]{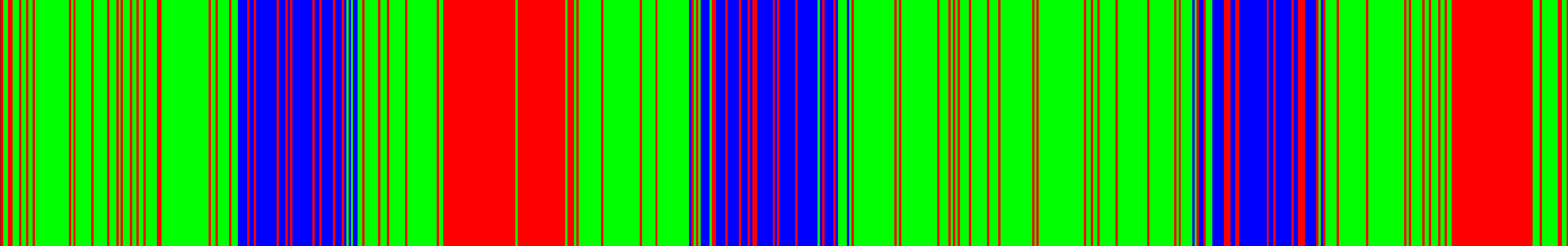}
\end{tabular}
\endgroup
	\centerline{\scriptsize{(d)}\includegraphics[width=0.96\linewidth,trim={3.9cm 0 3.2cm 0},clip]{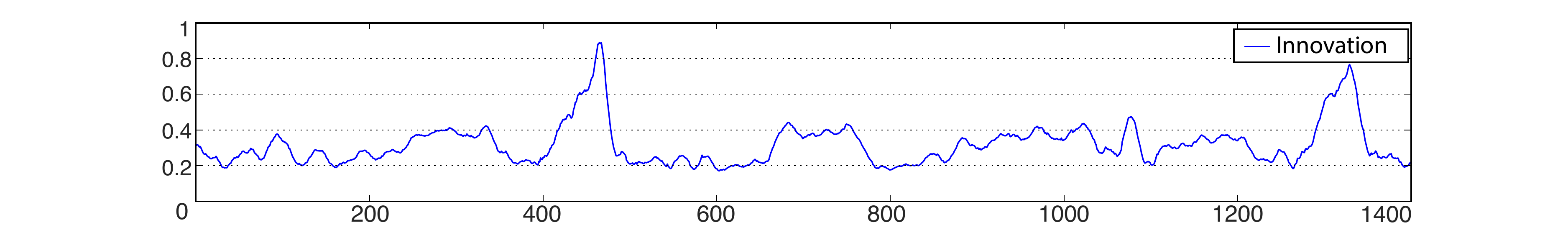}}
	\centerline{\scriptsize{(e)}\includegraphics[width=0.96\linewidth,trim={4.7cm 0 3.5cm 0},clip]{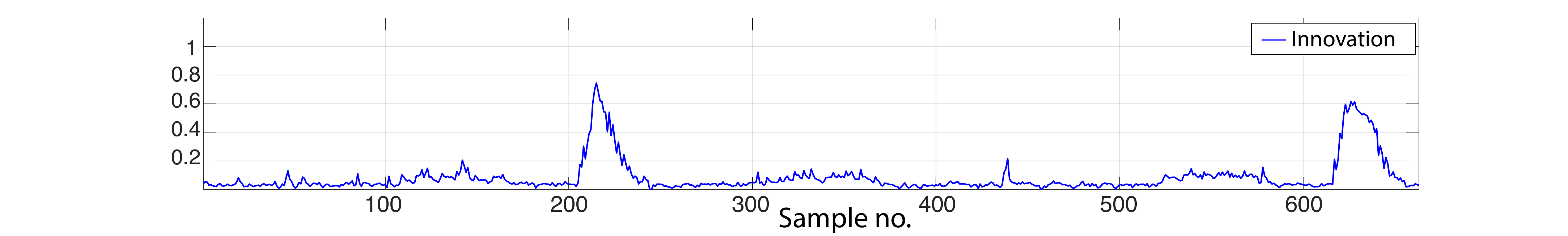}}

	\caption{Abnormality in the emergency stop scenario: (a) ground truth labels. (b) and (c) color-coded transition of states from PL and SL, respectively. (d) and (e) generated abnormality signal (innovation) $\{C_k^{m'}\}$ and $\{S_k^m\}$, respectively. The horizontal axis represents the sample number, and the vertical axis shows the innovation values (abnormality signal).}
	\label{fig:teststop}
\end{figure}
\noindent{\textbf{Emergency stop maneuver}:}  This scenario is shown in Fig. \ref{fig:subplans}-c, where the agent performs an emergency stop for a pedestrian to cross. Accordingly, Fig. \ref{fig:teststop} displays the results of abnormality detection for the highlighted time slice shown in Fig. \ref{fig:subplans}-c.   
In Fig. \ref{fig:teststop}-a, the red bars indicate the abnormality areas, i.e., where the vehicle stops and waits until the pedestrian crosses. Such anomaly areas are represented as dummy super states from \ac{pl} (light-blue color in Fig. \ref{fig:teststop}-b) with high scores in the abnormality signal (see Fig. \ref{fig:teststop}-d). The generated abnormality signal from \ac{pl} increases smoothly as the agent gets a better visual of the pedestrian and reaches the peak when the agent (vehicle) stops and has a full visual of the walker. Once the pedestrian passes and the agent starts to continue its linear path, the signal drops sharply. Similarly, the abnormality signal from the \ac{sl} representation model, see Fig. \ref{fig:teststop}-e, shows two peaks corresponding to high innovation values. Those peaks represent the abnormality patterns associated with the emergency stop maneuver. In contrast with the \ac{pl} signal, the \ac{sl} signal reaches a sharp peak and then smoothly goes back to the normal level. Such a pattern indicates that the vehicle stops immediately and waits for a while, then it increases the velocity to start moving again under the normal conditions. As a consequence, different motion patterns with respect to those predicted are detected using the innovation values. The color-coded super-states also confirm this; see Fig. \ref{fig:teststop}-c, where the green and dark-blue states are continued longer than what expected with respect to the normal pattern that learned from the previous observations.

\section{Discussion}
\label{sec:Discussion}

\noindent{\textbf{The cross-modal representation}: }One of the novelties in our paper is using \ac{gan}s for a {\em multi-channel} data representation. Specifically, we use appearance and motion (optical-flow) information: a two-channel approach that has been proved to be empirically significant in previous works. Moreover, we propose to use a cross-channel strategy, where we train two networks that transform raw-pixel images into optical-flow representations and vice-versa. The reasoning behind this relies on the architecture of our conditional generators $G$, which are based on an encoder-decoder (see Sec.~\ref{subsec:learn_PL}).  Such a channel-transformation task prevents $G$ from learning a trivial identity function by forcing $G$ and $D$ to construct sufficiently informative internal representations.

\noindent{\textbf{Private layer and shared layer cross-correlation}: }The PL and SL levels are providing complementary information regarding the situation awareness. As an instance, it has been observed that PL's super-states are invariant to the agent's location, while SL's super-states representation is sensitive to such spatial information. In other words, PL representation can be seen as the semantic feature of the agent's situation awareness (e.g., moving linearly, curving) regardless of its current location. Hence, the pattern of super-states sequences is repetitive; see how PL's super-states are repeated in Table \ref{tab:pri_ss}, e.g., Zone 1, 3, 5 are 4 correspond to the same super-states. In contrast, the SL representation includes spatial information, which generates more specific super-states for describing each zone, see table \ref{tab:pri_ss}. In light of the above, these two representations carry complementary information and define a cross-correlation between PL and SL situations. A coupled Bayesian network could represent such cross-correlation, enabling a potential improvement on the detection of abnormality and consequently boost the entire self-awareness model.

\noindent{\textbf{Performance of the method}:} Time and accuracy performances of the proposed method are shown in Table \ref{tab:performances}.
\begin{table}[!htb]
      \centering
        \begin{tabular}[\textwidth]{c c c c}
			\toprule
			N. particles&	time $(ms)$  &	$AUC_1(\%)$ &	$AUC_2(\%)$\\
			            &	[SL + PL]    &	            &	           \\
			\midrule
			5   &   7.6  + 190	& 91.0 & 85.1\\
			25  &   33.4 + 190	& 94.1 & 93.0\\
			50	&   65.4 + 190	& 96.2 & 96.1 \\
			\toprule
		\end{tabular}
		\caption{Computational time and accuracy of the proposed method. $AUC_1$ and $AUC_2$ are the area under the curve of the ROC in the U-turn and emergency stop respectively.}
		\label{tab:performances}
\end{table}

The computational time in Table \ref{tab:performances} refers to the time required to analyze, predict, and detect abnormalities in an instance of position and visual data. Increasing the number of particles in the SL improves the accuracy of the method while requiring more time. Since sampling times of positional are video data are $110\hspace{0.07cm} ms$ and $50 \hspace{0.07cm} ms$ respectively. SL's computational time with 50 particles ($\sim 65 \hspace{0.07cm} ms$) can handle real-time processing of positions. Nonetheless, PL's computational time ($\sim 190 \hspace{0.07cm} ms$) cannot handle real-time video processing in multiple GANs, however, this can be improved significantly using only a single model which either is capable to expand itself gradually \cite{ostapenko2019learning}, or learn the new concepts through a curriculum learning regime \cite{sangineto2018self}. All the experimental results reported in this work are obtained through a \ac{mjpf} that uses 50 particles.


\section{Conclusion and Future work}
\label{sec:con}
This work has presented a multi-modal approach to detect abnormalities in a moving vehicle. The proposed models consider two levels (\ac{sl} and \ac{pl}) that handle different types of information. \ac{sl} uses a state-space representation from an external observer, whereas \ac{pl} employs a state-space representation from the analyzed agent. \ac{sl} self-awareness employs a set of \ac{mkf}s coupled with a \ac{pf} to perform predictions. Innovations from filters are employed to build more complete models. On the other hand, \ac{pl} self-awareness is modeled as a hierarchy of cross-modal GANs that learn complex data distributions in a weakly-supervised manner. Scores of the Discriminator network are used to approximate the complexity of data. Namely, a set of distance maps between prediction and observation scores is used as a criterion for creating new hierarchical levels in the proposed GAN structure. Our approach facilitates incrementally learning complex data distributions. Experimental results on a ground vehicle show the capability of our methodology to recognize anomalies using multiple viewpoints, namely \ac{pl} and \ac{sl}. 
This work opens some future research paths, such as \textit{(i)} combining outputs from different sensorial sources that can lead to a more robust and unified \ac{sa} decision-making. \textit{(ii)} Optimizing of algorithms such that they can run on-device in a real-time fashion by using robotics middleware. \textit{(iii)} considering more complex scenarios that include the interaction with other artificial agents, e.g., another ground vehicle or a drone, to accomplish cooperative tasks.

\ifCLASSOPTIONcaptionsoff
  \newpage
\fi

\bibliographystyle{IEEEtran}
\bibliography{references}
\begin{IEEEbiography}[{\includegraphics[width=2.1cm,height=2.6cm,keepaspectratio]{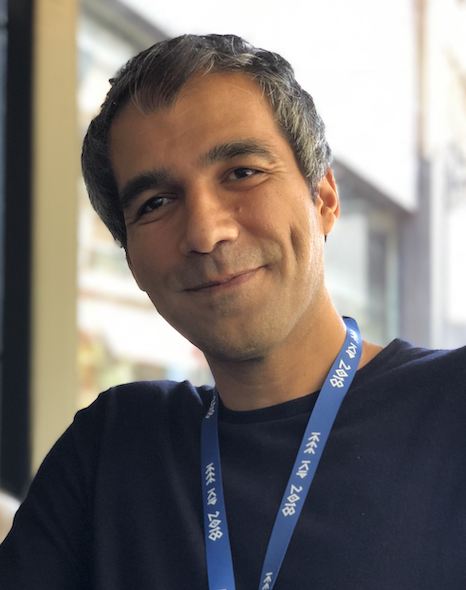}}]{Mahdyar Ravanbakhsh}
	He is a research fellow at Technische Universit\"at Berlin, Department of Electrical Engineering and Computer Science. He was a postdoctoral research fellow at University of Genoa, and a visiting researcher at the University of Trento.
	He obtained a PhD from the University of Genoa in 2019. His research lies at the intersection of machine learning and computer vision with an emphasis on deep learning with minimal supervision and/or limited data.
\end{IEEEbiography}

\begin{IEEEbiography}[{\includegraphics[width=2.1cm,height=2.6cm,keepaspectratio]{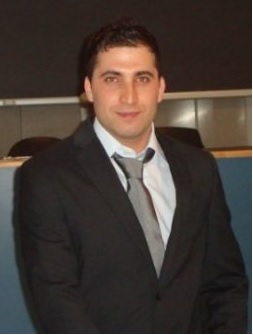}}]{Mohamad Baydoun}
	He obtained his titles of Telecommunication Engineer in 2011 and MS of Telecommunication Systems Engineering in 2015 from the University of Genova, Italy. Currently, he is a PhD candidate in interactive and cognitive environments at the University of Genova. His interests are machine learning and signal processing applications for abnormality detection purposes.
\end{IEEEbiography}

\begin{IEEEbiography}[{\includegraphics[width=2.1cm,height=2.6cm,clip,keepaspectratio]{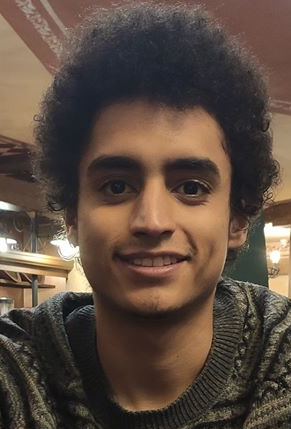}}]{Damian Campo}
	He received a degree in engineering physics from EAFIT University, Medellin, Colombia, in 2014, and a Ph.D. degree in cognitive environments from the University of Genoa, Italy, in 2018. Since 2018, he has been a postdoc researcher in the Department of Engineering and Naval architecture (DITEN), University of Genoa. His interests include the use of machine learning techniques and probabilistic theory for modeling and predicting the states of multisensorial data.
\end{IEEEbiography}

\begin{IEEEbiography}[{\includegraphics[width=2.1cm,height=2.6cm,clip,keepaspectratio]{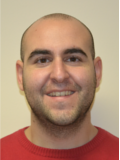}}]{Pablo Marin}
	He is graduated in Industrial Engineering: Industrial Electronics and obtain the degree of Industrial Electronics and Automation Engineering form Universidad Carlos III de Madrid in 2011, both in the same year. In 2013, he received the Master degree in Robotics and Automation in the same university. He is PhD candidate and assistant lecturer since 2013 and his current research interests include computer vision and autonomous ground vehicle.
\end{IEEEbiography}

\begin{IEEEbiography}[{\includegraphics[width=2.1cm,height=2.6cm,keepaspectratio]{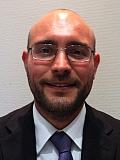}}]{David Martin}
	He is graduated in Industrial Physics (Automation) from the National University of Distance Education (UNED, 2002) and got his Ph.D. degree in Computer Science from the Spanish Council	for Scientific Research (CSIC) and UNED, Spain 2008. Currently, he is Professor and Post-Doc researcher at Carlos III University of Madrid and member of the Intelligent Systems Lab since 2011. In 2014, he was awarded with the VII Barreiros Foundation award to the best research in the automotive field. In 2015, the IEEE Society has awarded Dr. Martin as the best reviewer of the 18th IEEE International Conference on Intelligent Transportation Systems.
\end{IEEEbiography}
\vfill
\newpage 

\begin{IEEEbiography}[{\includegraphics[width=2.1cm,height=2.6cm,clip,keepaspectratio]{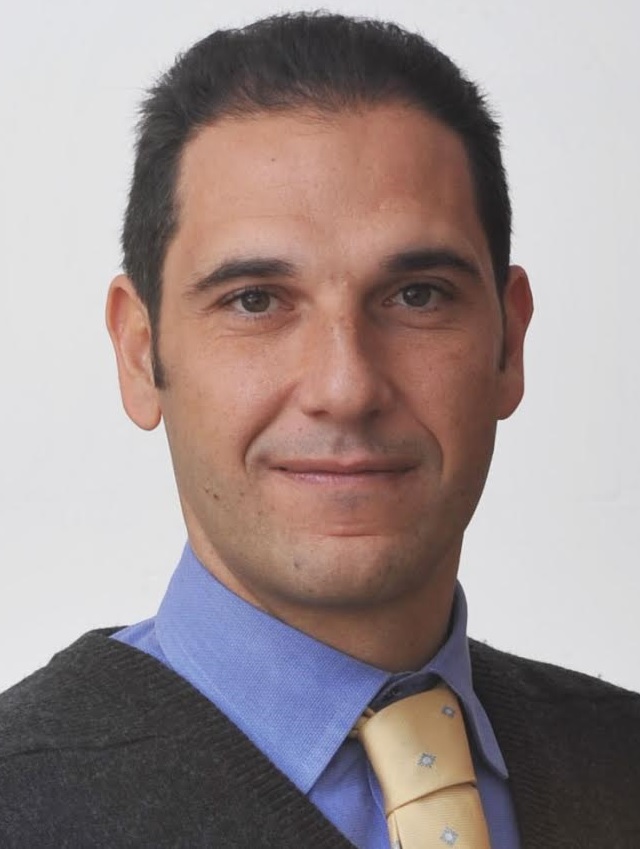}}]{Lucio Marcenaro}
	He enjoys over 15 years experience in image and video sequence analysis, and authored about 100 technical papers related to signal and video processing for computer vision. He is an Electronic engineer from Genoa University in 1999, he received his PhD in Computer Science and Electronic Engineering from the same
	University in 2003. From March 2011, he became Assistant Professor in Telecommunications for the Faculty of Engineering at the Department of Electrical, Electronic, Telecommunications Engineering and Naval Architecture (DITEN) at the University of Genova. 
\end{IEEEbiography}
\begin{IEEEbiography}[{\includegraphics[width=2.1cm,height=2.6cm,clip,keepaspectratio]{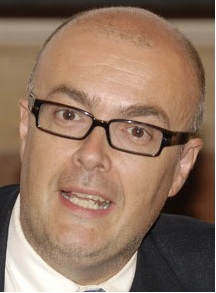}}]{Carlo Regazzoni}
	He is full professor of Cognitive Telecommunications Systems at DITEN, University of Genoa, Italy. He has been responsible of several national and EU funded research projects. He is currently the coordinator of international PhD courses on Interactive and Cognitive Environments involving several European universities. He served as general chair in several conferences and associate/guest editor in several international technical journals. He has served in many roles in governance bodies of IEEE SPS and He is serving as Vice President Conferences IEEE Signal Processing Society in 2015-2017.
\end{IEEEbiography}
\vfill

\end{document}